\documentclass[bibyear]{aa}

\usepackage{txfonts}
\usepackage{xcolor}
\usepackage{amsmath}
\usepackage{graphicx}
\usepackage{caption}
\usepackage{natbib}
\usepackage{pifont}
\usepackage{multirow}
\usepackage{soul}
\makeatletter
\renewcommand*\aa@pageof{, page \thepage{} of \pageref*{LastPage}}
\makeatother
\usepackage[colorlinks, linkcolor=teal, citecolor=olive]{hyperref}
\usepackage{orcidlink}

\setstcolor{red}

\newcommand\Msun{\mathrm{M}_\odot}
\newcommand\Rsun{\mathrm{R}_\odot}
\newcommand\Zsun{\mathrm{Z}_\odot}
\newcommand\cosmic{\texttt{COSMIC}}

\newcommand{\AGWB}{SGWB from extragalactic DWDs}
\newcommand{\Mchirp}{\mathcal{M}_\mathrm{c}}

\begin{document} 

\title{Gravitational-Wave Background from Extragalactic Double White Dwarfs for \textit{LISA}}

   \author{Guillaume Boileau\,\orcidlink{0000-0002-3576-6968}  \fnmsep\thanks{\email{guillaume.boileau@oca.eu}}
          \inst{1}
          \and
          Tristan Bruel\,\orcidlink{0000-0002-1789-7876}\inst{1,3,4}
          \and
          Alexandre Toubiana\,\orcidlink{0000-0002-2685-1538} \inst{3,4}
          \and
          Astrid Lamberts\,\orcidlink{0000-0001-8740-0127}\inst{1,2}
          \and
          Nelson Christensen\,\orcidlink{0000-0002-6870-4202} \inst{2}
          }

   \institute{Laboratoire Lagrange, Observatoire de la Côte d'Azur, Université Côte d'Azur, CNRS, France
         \and
             Artemis, Observatoire de la Côte d'Azur, Université Côte d'Azur, CNRS, CS 34229, F-06304 Nice Cedex 4, France
         \and
             Dipartimento di Fisica ``G. Occhialini'', Universit\`a degli Studi di Milano-Bicocca, Piazza della Scienza 3, 20126 Milano, Italy
         \and 
             INFN, Sezione di Milano-Bicocca, Piazza della Scienza 3, 20126 Milano, Italy
             }

   \date{Received September 15, 1996; accepted March 16, 1997}

 
\abstract
{Recent studies have revealed that the contribution of extragalactic double white dwarfs (DWD) to the astrophysical gravitational wave background could be detectable in the millihertz regime by the \textit{LISA} space mission. Conversely, the presence of this background could hamper the detection of cosmological backgrounds, which are one of the key targets of gravitational-wave astronomy.}
{We aim to confirm the amplitude and spectrum of the extragalactic DWD background and estimate its detectability with \textit{LISA} under different assumptions. We also aim at understanding the main uncertainties in the amplitude and frequency spectrum and estimate whether the signal could be anisotropic.}
{We use population synthesis code \cosmic~with several assumptions on binary evolution and initial conditions. We also incorporate a specific treatment to account for the episodes of mass transfer and tidal torques after formation of the DWDs.
}
{Our study is in global agreement with previous studies, although we find a lower contribution at high frequencies, due to a different treatment of mass transfer in stellar binaries. We find that the uncertainties in the amplitude are dominated by the star formation model, and to a lesser degree by the binary evolution model. The inclusion of tidal effects and mass transfer episodes in DWDs can change the amplitude of the estimated background up to a factor 3 at the highest frequencies. For all the models we consider, we find that this background would be easily detectable, with SNR values from~100 to more than 1000 by \textit{LISA} after 4 years of observations.
Under the hypothesis of an homogeneous Universe beyond 200 Mpc, anisotropies associated to the astrophysical population of DWDs will likely not be detectable. We provide phenomenogical fits of the background produced by extragalactic DWDs under different assumptions to be used by the community.
 }
{We demonstrate significant variability in gravitational wave background predictions, emphasizing uncertainties due to different astrophysical assumptions. We highlight the importance of determining the position of the knee in the gravitational wave background spectrum, as it provides insights on mass transfer models. The prediction of this background is of critical importance for \textit{LISA} in the context of observing other backgrounds.}


\maketitle
\section{Introduction}\label{sc:Intro}

The Laser Interferometer Space Antenna mission (\textit{LISA}), led by the European Space Agency (ESA), is designed to detect gravitational waves (GWs) in the frequency range $[1\times 10^{-4},1]$ Hz, with a projected launch in 2035 \citep{amaroseoane17, colpi24}. The diversity of sources detected by \textit{LISA} will greatly enhance our understanding of both the broader cosmological properties of the Universe and the structure of our Galaxy. Aside from individually resolved sources, \textit{LISA} is expected to observe several stochastic GW backgrounds (SGWB) which result from the cumulative effect of numerous individual sources that are too faint to be resolved separately~\citep{caprini18, christensen19}. This superposition produces a detectable background signal, quantified using the dimensionless parameter $\Omega_\mathrm{GW}$, which represents the energy density of the SGWB with respect to the closure density of the Universe. 

A guaranteed contribution to the SGWB comes from the superposition of unresolved compact binaries within our Galaxy, primarily double white dwarfs (DWDs). The confusion signal they create is known as the \textbf{Galactic foreground} ~\citep{ungarelli01, nissanke12, boileau21, karnesis21, toubiana24, buscicchio24, criswell24}. This foreground is challenging to analyze due to its observed non-stationarity caused by \textit{LISA}'s orbit in combination with the shape of the Milky Way. On the other hand, this annual modulation aids in its separation from the other, isotropic, backgrounds when analyzing \textit{LISA} data. 

An isotropic \textbf{cosmological SGWB} is anticipated to originate from a variety of processes in the early Universe \citep{caprini09, christensen19, mazumbar19, auclair23}, whose signals may combine into a single background. Detecting this cosmological SGWB is one of the key scientific objectives of \textit{LISA}, but such a signal must be carefully disentangled from other astrophysical backgrounds. Spectral separation techniques—such as parametric estimation—can help isolate individual components of the SGWB. In this context, Bayesian inference provides a model-dependent framework that combines prior knowledge with observational data to update the probability distributions of hypotheses or model parameters. For such methods to succeed, all non-cosmological contributions to the SGWB must be accurately characterised and parametrised.

For instance, an isotropic \textbf{astrophysical SGWB} is expected, due to the superposition of GWs emitted by inspiraling compact binaries throughout the Universe. Observations by the LIGO/Virgo/KAGRA collaboration (LVK) have provided an upper limit on the contribution to the astrophysical GW background coming from binaries containing black holes and/or neutron stars \citep{chen19, abbott21}.

Extragalactic DWDs are also expected to contribute to the astrophysical SGWB, but major uncertainties remain regarding their contribution, since this population of sources is not observable by any current facility. 
They were first introduced in \citet{schneider01}, and \citet{farmer03} later estimated the SGWB they produce, obtaining $\Omega_{\text{GW}}(1 {\rm mHz}) \in [1, 6] \times 10^{-12}$. 

Recently, \citet{staelens24} revisited this estimate, finding a 60\% higher amplitude than previous models, and concluding that DWDs likely dominate the SGWB in this frequency range.  The following study by \citet{hofman24} examined the impact of stellar metallicity, star formation rate density (SFRD), and binary evolution models, finding that the SGWB produced by extragalactic DWDs can vary by up to a factor 5, but consistently remains the primary source detectable by future detectors like \textit{LISA}, dominating the contribution from binaries made of black holes and/or neutron stars. Thus, the extragalactic DWD background appears to be a guaranteed \textit{LISA} source, which will impact the detectability of other sources, in particular the cosmological SGWB. 

This paper presents an alternative study of the modeling and understanding of the \AGWB. In the \textbf{Method} section (\S\ref{sc:Method}), we describe the \textit{LISA} context, the theoretical framework for SGWB modeling, the population synthesis of DWDs, and the development of a dedicated code to compute the \AGWB. The \textbf{Results} section (\S\ref{sc:Results}) starts with a validation of our model against previous studies, an analysis of how stellar evolution assumptions affect the resulting DWDs population, and investigations of the impact of taking into account matter interactions within DWDs on the SGWB and of potential anisotropies in the \AGWB\ signal. In the \textbf{Discussion} (\S\ref{sc:Discussion}) we explain the limits of our model and comment on the observability of the background.  The Conclusions sections (\S\ref{sc:Conclusion}) synthesizes our findings and suggests future research directions to deepen our understanding of the SGWB. In all our subsequent calculations, we adopt the cosmological parameters from the Planck 2018 results \citep{planck20}, with best-fit parameters $H_0 = 67.66 \, \text{km} \, \text{s}^{-1} \, \text{Mpc}^{-1} $, $ \Omega_m = 0.30966 $, and $ T_{\text{CMB}} = 2.7255 \, \text{K} $.

\section{Method}\label{sc:Method}

In this work, we estimate the contribution of extragalactic DWDs to the SGWB detectable by \textit{LISA}. Section~\ref{subsc:Context} sets the mission context, including instrument and observation assumptions. Section~\ref{subsc:Formalism} outlines the theoretical framework and SGWB energy density equations. In Section~\ref{subsc:Cosmic}, we present the population synthesis models for DWDs and Section~\ref{subsc:code} describes how these models are combined with various cosmic star formation histories to compute the \AGWB. 

\subsection{\textit{LISA} sensitivity to stochastic backgrounds}
\label{subsc:Context} 

We adopt the sensitivity curve corresponding to an equilateral triangle with stationary instrumental noise, as described in the \textit{LISA} red book \citep{colpi24, thorpe19}. This assumption allows us to directly model the SGWB signal without any need for additional instrumental corrections. 

It is standard practice to express the signal in terms of the GW energy density power spectrum, which can be related to the GW strain power spectral density. 
We introduce the energy density sensitivity \( \Omega_s(f) \), which is determined by the sensitivity of the detector's strain measurements $\Omega_{s}(f) = \frac{4\pi^2}{3H_0^2} f^3 S_n(f) = \frac{4\pi^2}{3H_0^2} f^3 N_X(f) /{\cal R}(f), $ where \( S_n(f) \) represents the noise power spectral density of the detector, \( {\cal R}(f)\) the \textit{LISA} response function \citep{babak2021} and \(N_X(f)\) the power spectral density of the \textit{LISA} noise \(X\) Channel \citep{tinto04}. 

For the rest of the paper we assume the equal correlation hypothesis and the \textit{LISA} noise model is given by the power-law integrated sensitivity \citep[PI curve, following][]{thrane13} for four years of observation time and a fixed value of signal-to-noise ratio SNR=10 (gray solid line on figures, see e.g. \citet{babak2021}).
The PI curve is a graphical tool used to represent the sensitivity of GW detectors to stochastic backgrounds characterized by power-law spectra. It indicates the minimum energy density of a stochastic background that a detector can observe across different spectral ranges, assuming the background is isotropic, stationary, and Gaussian.

While the PI curve is a valuable tool for assessing detector sensitivity to idealized SGWBs, its application to specific cases like \textit{LISA} presents limitations. \textit{LISA} is a unique space-based detector with arms forming an equilateral triangle and collocated interferometers, which implies distinct challenges compared to ground-based detectors. Moreover, certain astrophysical SGWBs, such as those from extragalactic white dwarf binaries, may deviate from the assumptions of the powerlaw spectrum, isotropy, stationarity, and Gaussianity. Consequently, more specialized analysis methods are necessary to evaluate the detectability of such backgrounds with \textit{LISA}. In particular, two key limitations arise when applying the PI approach to the \AGWB. First, the background is not stationary: the Galactic foreground contributing to the signal is modulated over the course of a year due to \textit{LISA}'s orbital motion. Second, the spectral shape of the \AGWB\ is not a simple power-law, as typically assumed in PI analyses, but results from a complex population synthesis shaped by astrophysical processes.
Consequently, more specialized analysis methods are necessary to evaluate the detectability of such backgrounds with \textit{LISA}. The PI curve has often been used as a standard framework in previous studies to evaluate \textit{LISA}'s sensitivity. In our case, we include it solely to enable comparison with earlier work. We do not interpret it further, as it does not reflect the physical complexity of the \AGWB.

The confusion noise generated by unresolved galactic DWD systems is modeled from \citep{robson19} as
\begin{equation}
 S_c(f) = A f^{7/3}_{\text{Hz}} e^{-f_{\text{Hz}} \alpha + \beta f_{\text{Hz}}\sin(\kappa f_{\text{Hz}}) } \left[1 + \tanh\left(\gamma(f_k - f_{\text{Hz}})\right)\right] ,
\end{equation}
where $f_{\text{Hz}} = f / (1\, \text{Hz})$, with the parameters $\alpha = 0.138$, $\beta = -221$, $\gamma = 1680$, and $f_k = 0.00113$. The corresponding dimensionless energy spectral density can be determined from $\Omega_{\text{GW},c}(f) = 4 \pi^2 S_c(f) f^3 / (3H_0^2)$.

Using the latest estimates from the LVK Collaboration, we model the total contribution of black hole binaries, neutron star binaries, and black hole-neutron star binaries to the SGWB as the normalised energy density $\Omega_{\text{NS/BH}}(f)$ such that:
\begin{equation}
    \Omega_{\text{NS/BH}}(f) = A_{\text{NS/BH}}(f/25\,\text{Hz})^{2/3} ,
\end{equation}
with $A_{\text{NS/BH}}=6.9^{+3.0}_{-2.1} \times 10^{-10}$ \citep{abbott23}.

\subsection{SGWB formalism and computation}
\label{subsc:Formalism}

The SGWB generated by compact binaries comes from the superposition of a large number of weak, independent, and unresolved sources. 
Its energy-density spectrum can be caracterised by the dimensionless quantity \citep{allen99}:
\begin{equation}
\label{eq:omega}
    \Omega_{\text{GW}}(f) = \frac{f}{\rho_c c^2}\frac{\mathrm{d}\rho_{\text{GW}}}{\mathrm{d}f},
\end{equation}
where \(\mathrm{d}\rho_{\text{GW}}\) is the energy density of this background signal within the frequency range $\left[\ f,f+\mathrm{d}f\ \right[$, and \(c\) is the speed of light. \(\rho_c = 3H_0^2/8\pi G\) is the critical energy density of the Universe, where \(H_0\) is the Hubble-Lemaitre constant and \(G\) the gravitational constant.

In a homogeneous, isotropic, and expanding Universe, the dimensionless energy spectral density $\Omega_{\text{GW}}$ measured today can be expressed as \citep{phinney01, regimbau08} :
\begin{equation}
\label{eq:omega_integral}
    \Omega_{\text{GW}}(f_r) = \frac{f_r}{\rho_c c^2} \int_0^{+\infty} N(z) \, \frac{1}{(1 + z)} \left. \frac{\mathrm{d}E_{\text{GW}}}{\mathrm{d}f_e} \right|_{f_e = f_r (1 + z)} \mathrm{d}z,
\end{equation}
where \(f_e = f_r (1+z) \) is the GW frequency of a source in its rest frame, and \(\mathrm{d}E_{\text{GW}}/\mathrm{d}f_e \) is its energy spectrum. The function \(N(z) \) denotes the number of such sources emitting GWs at frequency \(f_e\) per unit redshift and per unit comoving volume.

Following \citep{farmer03}, Eq.~\ref{eq:omega} can be rewritten considering a SGWB produced by binaries and in terms of specific flux development as:
\begin{equation}
    \Omega_{\mathrm{GW}}(f_\mathrm{r}) = \frac{1}{\rho_c c^3} f_\mathrm{r}F_{f_\mathrm{r}} ,
\label{eq:om_th}
\end{equation}
where $F_{f_\mathrm{r}}$ is the specific flux received in GWs today at frequency $f_\mathrm{r}$. In the case of a single source located at redshift $z$ and emitting GWs at a frequency $f_\mathrm{e}$ with a specific luminosity $L_{f_\mathrm{e}}$, this specific flux reads:
\begin{equation}
    F_{f_\mathrm{r}} = \frac{L_{f_\mathrm{e}}}{4\pi d_\mathrm{L}(z)^2} \left(\frac{\mathrm{d}f_\mathrm{e}}{\mathrm{d}f_\mathrm{r}}\right) ,
\end{equation}
where $d_\mathrm{L}(z)$ is the luminosity distance to the source located at redshift $z$. 
In reality, the \AGWB\ emanates from an immense number of sources located over a large redshift range, and $F_{f_\mathrm{r}}$ is the sum of the specific fluxes of all these individual systems. \citet{staelens24} find that the number of DWDs making up this GW background can be as high as $10^{17}$. As it is impossible to simulate all these sources and compute the actual discrete sum of all their contributions, we make the assumption that they are isotropically distributed over the sky and express the specific flux as:
\begin{align}
    \begin{split}
    F_{f_\mathrm{r}} 
    &= \int_{z=0}^{\infty} \frac{l_{f_\mathrm{e}}(z)}{4\pi d_\mathrm{L}(z)^2} \left(\frac{\mathrm{d}f_\mathrm{e}}{\mathrm{d}f_\mathrm{r}}\right)\mathrm{d}V(z)\\
    &= \int_{z=0}^{\infty} \frac{l_{f_\mathrm{e}}(z)}{(1+z)^2} \left(\frac{\mathrm{d}f_\mathrm{e}}{\mathrm{d}f_\mathrm{r}}\right)\mathrm{d}\chi(z) ,
    \end{split}
\label{eq:integral_Om_AGWB}
\end{align}
where $l_{f_\mathrm{e}}(z)$ is the specific luminosity density of sources at redshift $z$ emitting with a GW frequency $f_\mathrm{e}$ and $\mathrm{d}V(z)$ is the comoving volume element at redshift $z$. $\mathrm{d}\chi(z)$ is the comoving distance and $\mathrm{d}V(z)=4\pi d_\mathrm{M}(z)^2\mathrm{d}\chi(z)$ with $d_\mathrm{M}(z)$ the proper motion distance such that $d_\mathrm{L}(z)=(1+z)d_\mathrm{M}(z)$. This formula expresses the idea that the GW flux received in the present-day at a given frequency $f_\mathrm{r}$ can originate from sources at all redshifts, with various emitting frequencies $f_\mathrm{e}$ that are redshifted to reach us observers with the frequency $f_\mathrm{r}$ following $f_\mathrm{e}=(1+z)f_\mathrm{r}$.

We note here that, in this work, we do not take into consideration the contribution from any particular galaxy, such as the Milky Way or satellite galaxies~\citep{pozzoli25}. The computation of the SGWB is entirely based on the assumption of isotropic homogeneous Universe at all scales. We will see later (see \S\ref{subsc:Anisotropies}) that, even though this assumption does not hold at the lowest redshifts, the SGWB is almost entirely dominated by sources at redshifts where the Universe can be considered as such, and that the treatment of local galaxies would have only a very minor effect.

To estimate the energy-density spectrum of the \AGWB, we discretize the integral in Eq.~\ref{eq:integral_Om_AGWB}. We split the redshift range from $z=0$ to $z=8$ in 20 bins corresponding to a linear spacing in lookback time, and the received GW frequencies are logarithmically spaced in 50 bins between $10^{-5}\ \mathrm{Hz}$ 
and $10^{-1}\ \mathrm{Hz}$. For readability purposes, in all the figures showing $\Omega_\mathrm{GW}(f)$ we only plot half of the frequency bins in the range $[5\times10^{-5},5\times10^{-2}]\ \mathrm{Hz}$.
The total flux received in any frequency bin $\left[f_\mathrm{r1},f_\mathrm{r2}\right]$ can now be expressed in the discrete form:
\begin{align}
    \begin{split}
    F_{f_\mathrm{r1}\rightarrow f_\mathrm{r2}} &= \int_{f_\mathrm{r1}}^{f_\mathrm{r2}} F_{f_\mathrm{r}} \mathrm{d}f_\mathrm{r}\\
    &= \int_{f_\mathrm{r1}}^{f_\mathrm{r2}} \left(\sum_{z_\mathrm{i}} \frac{l_{f_\mathrm{e}}(z_\mathrm{i})}{(1+z_\mathrm{i})^2} \left(\frac{\mathrm{d}f_\mathrm{e}}{\mathrm{d}f_\mathrm{r}}\right)\Delta\chi(z_\mathrm{i})\right) \mathrm{d}f_\mathrm{r}\\
    &= \sum_{z_\mathrm{i}}\int_{f_\mathrm{r1}(1+z_\mathrm{i})}^{f_\mathrm{r2}(1+z_\mathrm{i})} \frac{l_{f_\mathrm{e}}(z_\mathrm{i})}{(1+z_\mathrm{i})^2} \mathrm{d}f_\mathrm{e}\Delta\chi(z_\mathrm{i}) .
    \end{split}
\label{eq:flux}
\end{align}

We now evaluate the specific luminosity density $l_{f_\mathrm{e}}$ in our different redshift bins and for each frequency bin $\left[f_\mathrm{r1},f_\mathrm{r2}\right]$ in order to compute the fluxes $F_{f_\mathrm{r1}\rightarrow f_\mathrm{r2}}$ and finally $\Omega_{\mathrm{GW}}$. We use catalogues of DWDs generated through population synthesis (described in further details in \S\ref{subsc:Cosmic}) and estimate the specific luminosity density as \citet{staelens24}:
\begin{equation}
    l_{f_\mathrm{e}}(z)=\sum_{k\in\mathrm{DWDs}}\frac{\mathrm{d}E_\mathrm{GW}}{\mathrm{d}f_\mathrm{e}}(k)\dot{f_\mathrm{e}}(k)\times n_k(f_\mathrm{e},z) ,
\label{eq:lfe}
\end{equation}
where $\mathrm{d}E_\mathrm{GW}/\mathrm{d}f_\mathrm{e}(k)$ is the energy spectrum of the DWD $k$ emitting GWs at frequency $f_\mathrm{e}$, $\dot{f_\mathrm{e}}(k)$ is its frequency evolution, and $n_k(f_\mathrm{e},z)$ is the specific number density of such system at redshift $z$.

The energy spectrum can be approximated by the leading quadrupole order in the GW emission \citep{hawking87}:
\begin{equation}
    \frac{\mathrm{d}E_\mathrm{GW}}{\mathrm{d}f_\mathrm{e}}(k)=\frac{\pi^{2/3}}{3}G^{2/3}\mathcal{M_\mathrm{c}}^{5/3}f_\mathrm{e}^{-1/3} ,
\end{equation}
where $\Mchirp$ is the chirp mass of the system $k$.
It is defined as $\mathcal{M}_\mathrm{c}=(m_1m_2)^{3/5}/(m_1+m_2)^{1/5}$, with $m_1$ and $m_2$ the two DWD masses.

The specific number density can be written,
\begin{equation}
    n_k(f_\mathrm{e},z) =\mathcal{R}_k(f_\mathrm{e},z)/\dot{f_\mathrm{e}} ,
\label{eq:nk}
\end{equation}
with $\mathcal{R}_k(f_\mathrm{e},z)$ the rate density at which systems $k$ start emitting GWs at a frequency $f_\mathrm{e}$ at redshift $z$. 
In the following subsections \S\ref{subsc:Cosmic} and \S\ref{subsc:code}, we present how we combine results from population synthesis with a cosmological star formation history to compute these factors $\mathcal{R}_k(f_\mathrm{e},z)$.

The dimensionless energy density can finally be expressed using this total flux of GWs received in the frequency bin $\left[f_\mathrm{r1},f_\mathrm{r2}\right]$ as:
\begin{equation}
    \Omega_{\mathrm{GW}}(\overline{f_{12}}) = \frac{1}{\rho_c c^3} \frac{\overline{f_{12}}\times F_{f_\mathrm{r1}\rightarrow f_\mathrm{r2}}}{f_\mathrm{r2}-f_\mathrm{r1}} ,
\end{equation}
where \(\overline{f_{12}} \) is the mean frequency of the frequency bin $\left[f_\mathrm{r1},f_\mathrm{r2}\right]$.

In order to assess the detectability of the \AGWB, we compute the signal-to-noise ratio (SNR) following \citep{smith19}:
\begin{equation}
\mathrm{SNR}^2 = T_\mathrm{obs} \int \frac{\Omega_{\mathrm{GW}}^2(f)}{\Omega_{\mathrm{sens}}^2(f)} \, \mathrm{d}f,\label{subs:snr}
\end{equation}
where $T_\mathrm{obs} = 4 $ and 10 years is the observation time and $\Omega_{\mathrm{sens}}(f)$ is the effective sensitivity that includes LISA's instrumental noise and the Galactic confusion noise:
\begin{equation}
\Omega_{\mathrm{sens}}(f) = \Omega_{s}(f) + \Omega_{\text{GW},c}(f).
\end{equation}

\subsection{Population synthesis of DWDs with \cosmic}
\label{subsc:Cosmic}

In this study, we use the rapid binary population synthesis code \cosmic\ \citep{breivik20} to simulate the evolution of isolated stellar binaries leading to the formation of DWDs. 
\cosmic\ is a code directly adapted from the Binary Stellar Evolution code \citep{hurley02}, with many additional prescriptions. The most relevant one for WD progenitors is the model of mass loss through stellar winds \citep[e.g.][]{vink01, vink05}. As described in \citet{hurley02}, the stability of mass transfer is determined in \cosmic\ using values of critical mass ratios $q_\mathrm{crit}=m_\mathrm{donor}/m_\mathrm{accretor}$ \citep[see e.g.][]{hjellming87} and the phase of common envelope (CE) evolution is modelled with the standard $\alpha\lambda$ energy formalism \citep[see e.g.][]{ivanova13}. In this formalism, $\lambda$ represents a binding energy factor of the envelope to the binary and the parameter $\alpha$ describes the efficiency with which orbital energy can be transferred to the envelope, eventually ejecting it if a sufficient amount of energy is passed on.

In our \texttt{default} set of population synthesis prescriptions, each stellar population that we simulate is initialised following the \citet{kroupa01} initial mass function (IMF) between $0.08$ and $150\ \Msun$. After individual stellar masses have been drawn from this distribution, each star is associated with a companion or not according to the mass-dependent binary fraction $f_\mathrm{b}(M)=\frac{1}{2}+\frac{1}{4}\mathrm{log}(M)$, for $0.08\leq M/\Msun \leq 100$ \citep{vanHaaften13}. We assume that all stars more massive than $100\ \Msun$ have a companion. Higher-order multiple systems are not considered here. Companion masses are sampled following a uniform distribution for the mass ratios $q=M_2/M_1$ \citep{goldberg94}. We use the distributions inferred in \citet{sana12} for orbital separations and eccentricities. To take into account the effect of metallicity on binary evolution and the formation of DWDs, we initialise and evolve stellar populations with 25 different metallicities, uniformly log-spaced such that the central values cover the range $[10^{-2},1]\ \Zsun$, where we assume $\Zsun=0.017$ after \citet{grevesse98}.

One particular feature of \cosmic\ is that the size of the total simulated population of stellar binaries can be set as a free parameter and determined over the course of each run. This allows \cosmic\ to continue initialising and evolving new binaries until the physical properties of the final population of DWDs have reached a certain convergence \citep[we refer the reader to][for details on the convergence criterion, that we have set to a value of -5 here]{breivik20}. As such, we warn the reader that the total number of DWDs obtained for one given run is not a representative quantity, but the total size of the simulated stellar population should also be taken into account when comparing results from different \cosmic\ models.

For the different prescriptions used to simulate stellar evolution and binary interactions, we follow, for the most part, the set of physical prescriptions proposed on the \cosmic\ \href{https://github.com/COSMIC-PopSynth/COSMIC/blob/master/examples/Params.ini}{GitHub page}. Regarding the phase of CE evolution, we consider a binding energy factor $\lambda$ that depends on stellar types following \citet{claeys14} and an efficiency parameter $\alpha=1$. 

\begin{table*}
    \centering
    \begin{tabular}{lc|ccc}
        \hline
        \hline
         & \texttt{default} & \texttt{fb1} & \texttt{multidim} & \texttt{$\alpha$4} \\
        \hline
        He-He & \(3.1\times10^4\) (12.6\%) & \(2.6\times10^4\) (13.9\%) & \(1.6\times10^4\) (9.1\%) & \(9.8\times10^4\) (21.7\%) \\
        \hline
        CO-He & \(1.41\times10^5\) (58.1\%) & \(1.07\times10^5\) (57.1\%) & \(1.04\times10^5\)  (58.3\%) & \(1.63\times10^5\)  (36.2\%) \\ 
        \hline
        CO-CO &\(6.4\times10^4\)  (26.3\%) & \(5.0\times10^4\)  (26.5\%) & \(4.9\times10^4\) (27.4\%) & \(1.82\times10^5\) (40.3\%) \\ 
        \hline
        ONe-He & \(2\times10^3\) (0.8\%) & \(1\times10^3\)  (0.7\%) & \(5\times10^3\)  (3.0\%) & \(1\times10^3\)  (0.2\%) \\ 
        \hline
        ONe-CO & \(3\times10^3\)  (1.3\%) & \(2\times10^3\)  (1.1\%) & \(3\times10^3\)  (1.8\%) & \(6\times10^3\)  (1.3\%) \\ 
        \hline
        ONe-ONe & \(2\times10^3\)  (0.8\%) & \(1\times10^3\)  (0.7\%) & \(\leq1\times10^3\)  (0.3\%) & \(1\times10^3\)  (0.2\%) \\ 
        \hline
        \textbf{Total} \boldmath$N_\mathrm{DWDs}$ & \(2.43\times10^5\) & \(1.87\times10^5\) & \(1.79\times10^5\) & \(4.51\times10^5\) \\ 
        \hline
        \hline
        \boldmath$M_\mathrm{SF}\ [\Msun]$ & $8.47\times10^7$ & $4.59\times10^7$ & $4.26\times10^7$ & $3.67\times10^7$ \\
        \hline
        \hline
    \end{tabular}
    \caption{Total number of each type of DWDs formed with our different \cosmic\ population synthesis models at solar metallicity, as well as the corresponding total mass of the stellar population simulated. Here we have already selected the systems that would have a redshifted GW frequency higher than $10^{-6}\ \mathrm{Hz}$ at $z=10$.}
    \label{table:COSMIC}
\end{table*}

Aside from out \texttt{default} model, we consider three additional models to measure the importance of the characteristics of the initial stellar population and of the prescriptions used to model their evolution. In \texttt{fb1}, we assume that all stars are born in binaries, i.e. that, regardless of the mass, the binary fraction is $f_\mathrm{b}=1$. This model serves as an extreme case for quantifying the uncertainty in the binary fraction during the astrophysical processes of star formation. In \texttt{multidim}, the initial binary parameters are no longer sampled sequentially, as in the \texttt{default} model described above, but all at once using the multidimensional distribution functions described in \citet{moe17}. Finally, in order to quantify some of the uncertainties during binary evolution and for the sake of comparison with \citet{staelens24, hofman24}, we also run a series of \cosmic\ simulations with a CE efficiency parameter $\alpha=4$ (in what follows, this model is referred to as \texttt{$\alpha$4}). A higher value of $\alpha$ means that energy is transferred more efficiently to the envelope. It is therefore ejected more easily and this typically results in post-CE systems with greater orbital separations, i.e~lower GW frequencies at formation.

As we are interested in the DWDs that contribute to the SGWB detectable with \textit{LISA}, and in order to reduce the size of the large populations of DWDs simulated with \cosmic, we apply a frequency threshold to select only the systems which, in the course of their evolution, have a chance of passing through the \textit{LISA} frequency band (0.1 mHz to 1.0 Hz with a GW strain spectral density \(10^{-21}\) to \(10^{-23}\), \citep{colpi24}). In practice, we select the systems that would have a redshifted GW frequency higher than $10^{-6} \mathrm{Hz}$ at redshift $z=10$.

For each WD in a binary, we look at the composition of its core and whether it is made of helium (He), carbon-oxygen (CO), or oxygen-neon (ONe). The number of each type of DWDs formed at solar metallicity with our different \cosmic\ set of models in are presented in Table~\ref{table:COSMIC}. 

The diversity in WD core composition arises from the dependence of stellar core evolution on the progenitor's initial mass. Stars with initial masses \(\lesssim 2.2 \Msun \) evolve into helium-core WDs, while intermediate-mass stars (\(\sim2.2 \)--\(\sim 6.5 \Msun \)) undergo helium and hydrogen shell burning eventually forming carbon-oxygen WDs. More massive stars (\(\gtrsim6.5 \Msun\)), depending on metallicity and convective over shooting, can ignite carbon and form oxygen-neon cores before envelope loss or collapse \citep{althaus10}.

It is important to note here that these numbers for DWDs do not account for any assumption on the cosmic star formation history or on the metallicity distribution across time in the Universe. They represent the catalogue of all the DWDs simulated with $\Zsun$ after applying the low-frequency cut-off. In our \texttt{default} model, the majority of systems are CO-He pairs ($\sim58\%$), followed by CO-CO ($\sim26\%$) and He-He pairs ($\sim13\%$). As they originate from more massive stars, the systems containing at least one ONe WD are naturally less common ($\sim3\%$ in total), with the rarest being the ONe-ONe pairs.

As the three models \texttt{default}, \texttt{fb1}, and \texttt{multidim}, use the exact same prescriptions to describe binary evolution but differ only in the properties of their initial stellar populations, they predict comparable relative numbers of each type of DWDs. On the other hand, the \cosmic\ model \texttt{$\alpha$4} using a higher CE efficiency parameter results in an overall larger efficiency for forming DWDs in the considered frequency range, and in notable differences in the relative number of types of DWDs.
This can be explained by the fact that, with a higher efficiency at ejecting the envelope, more systems can survive this phase of unstable mass transfer. More He-He systems are formed in this case.

More details on the physical properties of the DWDs at the time of their formation and on the formation efficiency between our different \cosmic\ models are presented in appendix \S\ref{app:DWDs}.

\subsection{Procedure for estimating the \AGWB}
\label{subsc:code}

\subsubsection{Metallicity-specific SFRD}
\label{subsubsc:SFRD}
In order to estimate the rate density factors $\mathcal{R}_k(f_\mathrm{e},z)$ in Eq.~\ref{eq:nk} at different redshifts and for different emitting frequencies, we relate the formation of DWDs to the formation of stars in the Universe. Throughout this study, we make the assumption that the metallicity-specific SFRD can be split in two different independent factors:
\begin{align}
    \begin{split}
    \frac{\mathrm{d}^3M_\mathrm{SFR}}{\mathrm{d}t\mathrm{d}V\mathrm{d}Z}(Z,z)&=\frac{\mathrm{d}^2M_\mathrm{SFR}}{\mathrm{d}t\mathrm{d}V}(z)\times \frac{\mathrm{d}P}{\mathrm{d}Z}(Z,z)\\
   &=\psi(z)\times \frac{\mathrm{d}P}{\mathrm{d}Z}(Z,z) ,
   \end{split}
\label{eq:ZSSFR}
\end{align}
where $\psi(z)$ is the cosmological SFRD as a function of redshift and $\mathrm{d}P/\mathrm{d}Z$ indicates the metallicity distribution function.

We use here a single model for the metallicity distribution function $\mathrm{d}P/\mathrm{d}Z$, taken as the \textit{preferred model} presented in \citet{neijssel19}. 
To take account of certain observational uncertainties in the SFRD, particularly at high redshifts, we use here three different models: the classic SFRD from \citet{madau14}, an updated empirical determination from \citet{madau17}, and a parametric form in time from \citet{strolger04}.
Following \citet{madau14} and \citet{madau17}, the SFRD as a function of redshift can be expressed as:
\begin{equation}\label{eq:Madau}
    \psi(z) = a \frac{(1 + z)^b}{1 + [(1 + z)/c]^d}\ \Msun\mathrm{yr}^{-1}\mathrm{Mpc}^{-3}.
\end{equation}
Alternatively, \citet{strolger04} parametrise the SFRD as a function of cosmic time (expressed in Gyr) as:
\begin{equation}\label{eq:strolger}
\widetilde{\psi}(t) = \tilde{a} \left(t^{\tilde{b}} e^{-t/\tilde{c}} + \tilde{d} e^{\tilde{d} (t - t_0)/\tilde{c}}\right)\, \Msun\mathrm{yr}^{-1}\mathrm{Mpc}^{-3} ,
\end{equation}
where $t_0$ is the current age of the universe (set in their fit at 13.47 Gyr).

Combining the results of our population synthesis of DWDs carried out using \cosmic\ (\S\ref{subsc:Cosmic}) with the models of metallicity-dependent star formation presented here, we can now estimate the rate density at which systems $k$ start emitting GWs at a frequency $f_\mathrm{e}$ at redshift $z$ as:
\begin{equation}
    \mathcal{R}_k(f_\mathrm{e},z)=\frac{\psi(z+\Delta z)}{M_\mathrm{SF}(Z_k)}\times\frac{\mathrm{d}P}{\mathrm{d}Z}(Z_k,z+\Delta z) ,
\label{eq:Rk}
\end{equation}
where $\Delta z$ indicates the variation of redshift corresponding to the time elapsed from the formation of the progenitor stellar binaries until the remnant DWD $k$ has evolved to emit GWs at frequency $f_\mathrm{e}$ at redshift $z$, and $M_\mathrm{SF}(Z)$ is the total initial mass of the stellar population simulated with \cosmic\ at metallicity $Z$.

 More details on these models of SFRD and on the metallicity distribution function are presented in appendix \S\ref{app:SFR}.  The best fit parameters for Eqs.~\ref{eq:Madau}-\ref{eq:strolger} can be found in Table~\ref{tab:sfr_models}. In ~\ref{app:metallicity} we also show that metallicity has a limited impact on DWD formation.

\subsubsection{Orbital evolution with GW emission}
\label{subsubsc:OrbitalEvol}

In the Newtonian limit, the orbital frequency of a binary system is given by:
\begin{equation}
    f_\mathrm{orb} = \frac{1}{2\pi}\sqrt{\frac{G (M_1 + M_2)}{a^3}},
\end{equation}
where $a$ is the semi-major axis of the orbit. 
For quasi-circular systems, the GW emission frequency of a binary relates to its orbital frequency as $f_{e} = 2\times f_{\mathrm{orb}}$.

Under the simplified assumption that the binary evolves purely by GW emission, the frequency evolution follows~\citep{farmer03}:
\begin{equation}
    \dot{f_\mathrm{orb}} = K f_\mathrm{orb}^{11/3} ,
\end{equation}
where 
\begin{equation}
\label{eq:fdotorb}
    K=\frac{96}{5}(2\pi)^{8/3}\left(\frac{G\mathcal{M}_\mathrm{c}}{c^3}\right) .
\end{equation}

By default, we consider that when one of the two WDs fills its Roche lobe, meaning that its radius extends beyond the surface inside which matter is gravitationally bound to it, the DWD merges and the system does not contribute to GW emission anymore \citep{lamberts19}. As such, we determine a maximum orbital frequency, and therefore a maximum GW emission frequency $f_\mathrm{e,max}$, for each system in our catalog. We discuss in the following subsection (\S\ref{subsc:Interaction} an alternative treatment taking into account the interaction between WDs.  

To compute the Roche lobe effective radius for the WD identified by the index $1$, we use the approximation from \citet{eggleton83}:
\begin{equation}
    R_\mathrm{L,1}=\frac{0.49q^{2/3}}{0.6q^{2/3}+\mathrm{ln}(1+q^{1/3})}\ a ,
\end{equation}
where $0<q=M_1/M_2<\infty$ and $a$ is the orbital separation of the binary. The minimum orbital separation that a binary can reach before one of its two components fills its Roche lobe is then given by: $a=\mathrm{max\left(R_\mathrm{WD,1}/R_\mathrm{L,1},\ R_\mathrm{WD,2}/R_\mathrm{L,2} \right)}$, where $R_\mathrm{WD,1}$ and $R_\mathrm{WD,2}$ are the radii of the two WDs.

From Eq.~\ref{eq:fdotorb}, we can then derive the time elapsed from the formation of a DWD until it reaches a given GW frequency \(f_\mathrm{e}<f_\mathrm{e,max}\) as:
\begin{equation}
    \Delta t = 2^{8/3}\frac{3}{8K}\left(f_\mathrm{e,0}^{-8/3} - f_\mathrm{e}^{-8/3}\right) ,
\end{equation}
where \(f_\mathrm{e,0}\) is the binary initial GW frequency.

In the case of evolution through GW emission only, we combine for each system the time delay from progenitor stellar binary to DWD formation predicted by \cosmic\ with the above expression of $\Delta t$ to estimate in Eq.~\ref{eq:Rk} the equivalent $z+\Delta z$ for different GW frequencies and different redshifts. 

For the systems that can never emit GWs at frequency $f_\mathrm{e}$ at redshift $z$, because it would take too long to reach this frequency, or they would merge before, or they form already at a higher frequency, we take $\mathcal{R}(f_\mathrm{e},z)=0$. 

\subsubsection{Treatment of interacting DWDs}
\label{subsc:Interaction}

In the case where we consider mass transfer and tidal effects during the evolution of DWD systems, the chirp mass is no longer constant and both $\Mchirp(t)$ and $f_\mathrm{e}(t)$ do not have analytical expressions. We evolve each DWD using the model of~\citet{toubiana24} and recover the simulated evolutions of $\Mchirp(t)$, $f_\mathrm{e}(t)$, and $\dot{f_\mathrm{e}}(t)$. 

In short, when the two WDs are close enough together, the lighter one, which is also the largest one, may overfill its Roche lobe and start transferring mass onto the heavier one. An intense mass-transfer episode occurs, which may cause a prompt merger. In our model, this occurs if the mass-transfer rate exceeds $10^{-2}\ \mathrm{M}_{\odot}\mathrm{yr}^{-1}$. Binaries that survive this mass transfer start out-spiralling. In some rare cases, the donor can successively overfill and underfill its Roche lobe, causing the orbital separation to oscillate, until the binary definitely out-spirals.

To compute the integral of Eq.~\ref{eq:flux}, we perform a change of variable as $f_\mathrm{e}\rightarrow t$ such that:
\begin{align}
    \begin{split}
    \int_{f_\mathrm{e}} l_{f_\mathrm{e}}(z)\mathrm{d}f_\mathrm{e} &= \frac{\pi^{2/3}}{3}G^{2/3}\int_{f_\mathrm{e}}\sum_{k}\Mchirp^{5/3}f_\mathrm{e}^{-1/3}\dot{f_\mathrm{e}} \mathcal{R}_k(f_\mathrm{e},z)\frac{\mathrm{d}f_\mathrm{e}}{\dot{f_\mathrm{e}}} \\
    &= \frac{\pi^{2/3}}{3}G^{2/3}\sum_{k}\mathcal{R}_k\int_{t}\Mchirp(t)^{5/3}f_\mathrm{e}(t)^{-1/3}\dot{f_\mathrm{e}}(t)\ \mathrm{d}t .
    \end{split}
\label{eq:realistic}
\end{align}

Binaries that out-spiral may contribute multiple times to the same frequency bins. To account for that effect, we use our numerical solution of $f_\mathrm{e}(t)$ to identify each occurrence of the binary crossing a given frequency bin and then sum the contribution of each of those passages
as the integral in Eq.~\ref{eq:realistic}. 
Each of the contributions is weighted by the corresponding factor defined in Eq.~\ref{eq:Rk}, where $\Delta z$ is the redshift variation corresponding to the time elapsed between the formation of the progenitor stellar binaries and that passage of the binary in the frequency bin. 

Unless otherwise specified, we always use the first simplified case throughout this paper and assume that DWD evolution is driven by GWs only, neglecting all the potential episodes of mass transfer and tidal interactions. The effects of this more realistic treatment are explored in \S\ref{subsc:tides}.

\section{Results}\label{sc:Results}

\subsection{Predictions for the \AGWB}
\label{subsc:Validation}

\begin{figure*}
    \centering
    \includegraphics{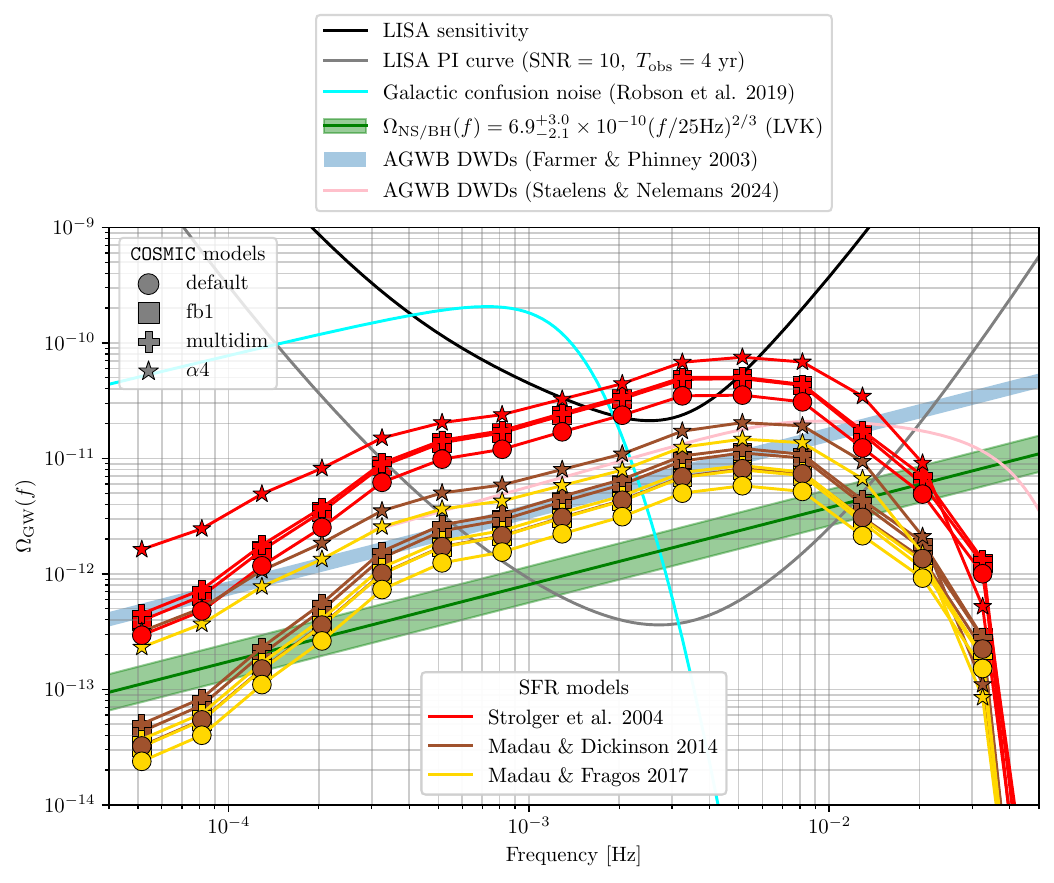}
    \caption{Spectral energy density of the \AGWB\ for various population synthesis models (see \S\ref{subsc:Cosmic}) and various SFRD models (see \S\ref{subsubsc:SFRD}). The different population synthesis models are indicated with the same markers as in Fig.~\ref{fig:etaDWD}. Dotted, solid, and dashed lines correspond to the SFR models from \citet{strolger04}, \citet{madau14}, and \citet{madau17} respectively.
    The black line represents the \textit{LISA} sensitivity curve \citep{colpi24}, while the gray one shows the PI curve \citep{thrane13} for four years of observation time and SNR=10. The light blue curve is the confusion noise from Galactic DWDs \citep{robson19} and the green shaded band is the 90\% credible interval of the SGWB from all compact object binaries estimated from observations of GWs \citep{abbott23}. As a comparison, the blue shaded band shows the \AGWB\ estimated in \citep{farmer03} and the pink curve indicates the the best-fit model obtained in \citep{staelens24}. In Appendix~\ref{app:parameters}, we discuss and summarize the fits of \AGWB\ of all SFR models and stellar population synthesis models as presented in Table~\ref{tab:appendix_parameters}.
    }
    \label{fig:omega_SFRs_cosmicmodels}
\end{figure*}

In Fig.~\ref{fig:omega_SFRs_cosmicmodels}, we show the spectral energy density obtained with our four different \cosmic\ models (see Table~\ref{table:COSMIC}) using our default SFRD model \citep{madau14}. We find consistent predictions across our different population synthesis models, although not exactly similar.
In particular, with a fixed SFRD model, the SGWB obtained with the \cosmic\ model \texttt{alpha4} is 2 to 4 times higher than the other three models in the frequency band $[5\times10^{-4},5\times10^{-3}]\ \mathrm{Hz}$. This difference is a direct consequence of the higher efficiency at forming DWDs with this model (see Fig.~\ref{fig:etaDWD}). With a similar approach and using different models of CE evolution, \citet{hofman24} report a factor of $\sim2$ between their highest and lowest estimates of the \AGWB\ at a fiducial frequency of 1 mHz.

The spread arising from the different population synthesis models is fairly similar across the different SFRD models and remains consistent with our previous description using the default SFRD model from \citet{madau14}. It is clear that the choice of SFRD model has a major impact on the global amplitude of the \AGWB. Indeed, as the fraction of stellar binaries that evolve to become DWDs is roughly constant over a large range of metallicities (see Fig.~\ref{fig:etaDWD}), the amplitude of the \AGWB\ naturally relates to the total mass of stars that have formed in the entire history of the Universe, or the integrated SFR. As the SFRD model from \citet{strolger04} predicts notably more star formation at $z>2$ than the two other SFRD models (see the left-hand panel of Fig.~\ref{fig:SFRs}), this naturally translates into much higher SGWB amplitudes. 
We find that the use of different SFRD models dominate the uncertainty in the SGWB amplitude, with a factor of up to 6-7, while different models of initial stellar populations and of binary evolution have a more moderate impact, with a factor 2-3 variation in the global amplitude of the predicted SGWB. This variation highlights the strong sensitivity of the \AGWB\ predictions to assumptions about the common-envelope phase and the global star formation rate.

In addition to modeling uncertainties, observational limitations must be considered. The Galactic foreground dominates the sub-mHz regime and decreases steeply above $\sim$2 mHz, while the extragalactic signal becomes most prominent between $\sim$3 and 7 mHz. Although the two components occupy partially distinct frequency ranges, their overlap at intermediate frequencies can still challenge signal separation. Extending the observation time improves sensitivity, but does not fully resolve the ambiguity between overlapping astrophysical contributions without additional constraints or targeted analysis techniques.

We compare our results with previous studies \citep[blue shaded area and pink curve, corresponding to][respectively]{farmer03,staelens24} and find a consistent overall picture. In particular, \citet{staelens24} use the same SFRD model from \citet{madau14} and simulate DWD formation with the population synthesis code \texttt{SeBa} \citep{portegies96,nelemans01,toonen12}. As we have used different codes and different physical prescriptions for certain treatments of binary interactions, it does not come as a surprise that we do not find an exact match to their predictions.
However, one very notable observation is the difference in the location of the frequency cut-off. While \citet{staelens24} find a decrease in the energy density spectrum for frequencies above $f_\mathrm{c}\sim 20\ \mathrm{mHz}$, all our models predict a sharp drop-off at a lower frequency $f_\mathrm{c}\sim7\ \mathrm{mHz}$. We explore in detail the origin of this discrepancy at high frequencies in the Discussion section (see Section~\ref{subsc:SeBa}).

These results underscore the significant importance of the assumptions made on star formation, stellar evolution, and binary interactions to predict a SGWB and their relevance for future GW observations. Apart from its global amplitude, we find no other distinguishable effect of the SFRD model on the \AGWB. For comparison, we also report in green the astrophysical SGWB expected from compact binaries composed of black holes and/or neutron stars, based on the LVK observations. In all our models, the contribution from extragalactic DWDs dominates in the \textit{LISA} range.

Our predictions for the SGWB of extragalactic DWDs come very close to the \textit{LISA} sensitivity curve (black curve) in the frequency band $[2\times10^{-3},8\times10^{-3}]\ \mathrm{Hz}$, and even distinctly above for the most optimistic \cosmic\ model \texttt{alpha4}. This suggests that the \textit{LISA} mission may allow us to put some strong observational constraints on the amplitude of this SGWB in this frequency range.

\begin{figure}
    \centering
    \includegraphics{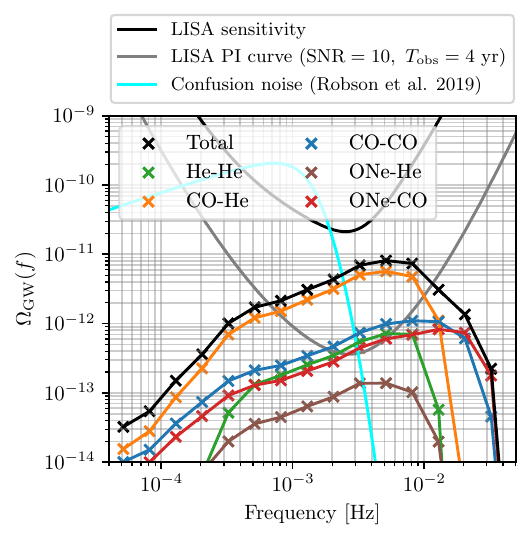}
    \caption{Spectral energy density of the \AGWB\ for the various types of DWDs, with our \texttt{default} \cosmic\ model and the SFRD from~\citet{madau14}. Different colours indicate the contribution from different type of DWD binaries. We do not show the contribution of ONe-ONe binaries here, as it is several orders of magnitude below due to their rarity. The sum of all the contributions is shown in black.
    Additional curves represent the \textit{LISA} sensitivity, the PI curve, and the confusion noise from Galactic DWDs.
    }
    \label{fig:omega_per_type}
\end{figure}

\begin{figure}
    \centering
    \includegraphics{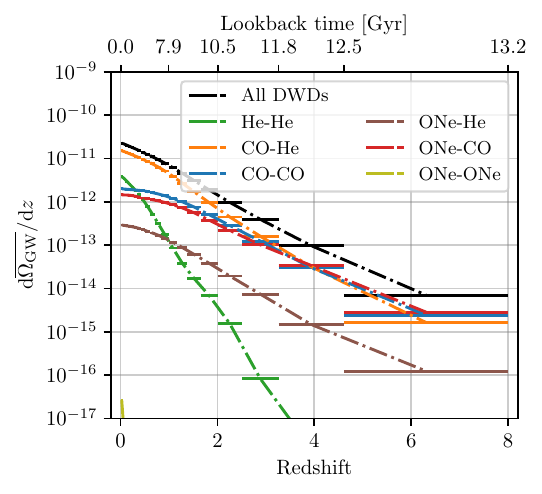}
    \caption{Spectral energy density integrated over frequency for different redshift bins. The total contribution from all redshift bins is shown in black. Different colours indicate the contribution from different types of DWD binaries.
    } 
    \label{fig:domega_per_z}
\end{figure}

\begin{table}[ht]
\centering
\caption{Signal-to-noise ratios (SNR) computed for each combination of SFR and population synthesis model. The integration follows \citet{smith19}, assuming a \textit{LISA} observation duration $T$.}
\label{tab:appendix_snr}
    \begin{tabular}{llcc}
    \hline
    \textbf{} & \textbf{SNR} & \textbf{$T=4$~yr} & \textbf{$T=10$~yr} \\
    \hline \hline
    \multicolumn{4}{c}{\textbf{Strolger}} \\ \hline
    \multirow{4}{*}{\rotatebox{90}{\shortstack{\textbf{Stellar} \\ \textbf{Model}}}}
    & \texttt{default}   & 748  & 1182 \\
    & \texttt{binfrac1}  & 1037 & 1640 \\
    & \texttt{multidim}  & 1076 & 1701 \\
    & \texttt{alpha4}    & 1493 & 2361 \\ \hline
    \multicolumn{4}{c}{\textbf{Madau Dickinson}} \\ \hline
    \multirow{4}{*}{\rotatebox{90}{\shortstack{\textbf{Stellar} \\ \textbf{Model}}}}
    & \texttt{default}   & 155  & 244 \\
    & \texttt{binfrac1}  & 212  & 335 \\
    & \texttt{multidim}  & 233  & 369 \\
    & \texttt{alpha4}    & 386  & 610 \\ \hline
    \multicolumn{4}{c}{\textbf{Madau Fragos}} \\ \hline
    \multirow{4}{*}{\rotatebox{90}{\shortstack{\textbf{Stellar} \\ \textbf{Model}}}}
    & \texttt{default}   & 111  & 176 \\
    & \texttt{binfrac1}  & 152  & 241 \\
    & \texttt{multidim}  & 167  & 264 \\
    & \texttt{alpha4}    & 279  & 441 \\ \hline
    \end{tabular}
\end{table}

Table~\ref{tab:appendix_snr} shows that the computed SNR varies by more than an order of magnitude across models, from \(\sim 100\) for conservative scenarios (e.g., \texttt{default} + \texttt{Madau Fragos}) up to nearly \(\sim 1500\) (e.g., \texttt{alpha4} + \texttt{Strolger}) for a 4-year mission, and reaching \(\sim 2400\) for a 10-year observation in the most favorable case. 

Even scenarios involving the \texttt{default} model paired with the \citet{madau14} or \citet{madau17} SFR yield SNR values well above unity ranging from \(\sim 100\) to \(\sim 400 \) for $T=4$~yr and up to \(\sim 600 \) for a 10-year mission. These very high SNR values demonstrate that , in all scenarios, the detection of the \AGWB\ by LISA is virtually guaranteed (see Table~\ref{tab:appendix_snr}), regardless of Galactic foreground or instrumental sensitivity limits. However, the degeneracies between the effect of the cosmological SFRD, the properties of initial stellar populations, and the nature of common envelope evolution will remain challenging to disentangle, as detection alone will not suffice to constrain these physical processes independently. These fundamental uncertainties in binary evolution modeling and star formation histories will affect the interpretation and parameter inference of the observed background. Therefore, while extending the observation period further increases SNR, the focus should shift toward accurately modeling these astrophysical uncertainties to fully exploit the detection, notably the position of the typical frequency at which the signal starts to drop, referred to here as the knee, which will have an impact on the astrophysical models.
Furthermore, if the SFRD turns out to be larger at high redshifts than the commonly used model of \citet{madau14}, then the \AGWB\ could have an even stronger impact on the effective sensitivity curve of LISA, particularly in the frequency band $[10^{-3},10^{-2}]\ \mathrm{Hz}$.

\subsection{Contribution of different DWD types to the SGWB}
\label{subsc:types}

Fig.~\ref{fig:omega_per_type} shows the energy density spectrum for the various types of DWDs, using our \texttt{default} \cosmic\ model and the SFRD from~\citet{madau14}. Across most of the frequency range the SGWB is dominated by CO-He sources. This is naturally expected from the high formation efficiency of this type of DWDs compared to the other types \citep[see Table~\ref{table:COSMIC} and e.g.][]{lamberts19}. 

At higher frequencies ($f_\mathrm{GW,r}\geq10^{-2}\ \mathrm{Hz}$), the contribution of the CO-He binaries becomes negligible and the background is made up of signals from CO-CO and ONe-CO binaries. This different behaviour arises from the fact that, being more massive, the WDs in these two types of systems are also smaller in radius \citep[see e.g.][for detail on the WD mass-radius relation]{kawaler97} and can therefore evolve to closer orbital separations before filling their Roche lobe.
We emphasise that this high-frequency region of the \AGWB\ is the most uncertain in our modelling, and that it is highly dependent on the treatment of binary stellar interactions. We explore these uncertainties in further detail in \S\ref{subsc:SeBa}.

The negligible contribution of ONe-ONe binaries originates from their rarity, as only $\sim0.7\%$ of all the DWDs are ONe-ONe in our \texttt{default} \cosmic\ model, but also from the fact that they form with typically very large orbital separations (see Fig.~\ref{fig:Dist_Freq_TypeDWD}) and that, for most of them, they can never enter the frequency band relevant to our analysis. 

We show in Fig.~\ref{fig:domega_per_z} the distribution of the spectral energy density integrated over all frequencies \(\overline{\Omega_\mathrm{GW}} = \int \Omega_\mathrm{GW}(f) \mathrm{d}\ln f\), in different redshift bins. Note that the vertical axis actually shows \(\mathrm{d} \overline{\Omega_\mathrm{GW}} / \mathrm{d}z \), i.e., the contribution normalised per size of the redshift bin. Looking at the different orders of magnitude of this integrated spectral energy density over the different redshift bins, it appears clear that most of the \AGWB\ arises from sources located at $z\leq1$. This result is not surprising in itself, as this redshift range already represents more than half of the cosmic evolution of the Universe.
Consistent with the results presented in Fig.~\ref{fig:omega_per_type}, the CO-CO binaries dominate the energy density in this relevant redshift range. 

However, it is interesting to note that at higher redshifts, CO-CO and ONe-CO binaries produce a higher SGWB than CO-He binaries. 
This is due to the typically long delays elapsing between the formation of their progenitor stellar binaries and that of the latter type DWDs. While CO-CO and ONe-CO binaries can be produced after only a few hundreds Myr of binary evolution \citep{istrate14}, the first CO-CO binaries take around $\lesssim 1$ Gyr to form. Consequently, in the first few billion years after the onset of star formation, DWDs were mostly CO-CO and ONe-CO, which explains why they dominate the SGWB.
The same argument applies to interpret the fact that the contribution of He-He binaries starts only after $z\sim4$. Indeed, stellar evolution and binary interactions prior to the formation of a He-He system typically extend over more than a billion years \citep[see e.g.][]{lamberts19}.

On the other hand, the negligible contribution of ONe-ONe binaries manifest only at very low redshifts due to their typically high initial orbital separations that translate into extremely long evolution time to evolve up to the \textit{LISA} frequency band. We show further results on the contribution of the different types of DWDs in appendix \S\ref{app:metallicity}.

\subsection{Including tidal effects and mass transfer during DWD evolution}
\label{subsc:tides}

We now explore the impact of interactions between DWDs, such as mass transfer and tidal interactions, on our predictions for the SGWB. In this sub-section, we no longer use the simplified case to describe the orbital evolution of all DWDs through only GW emission, but evolve each system independently using the model of \citet{toubiana24}, as described in \ref{subsc:Interaction}, and recover more realistic descriptions of their complete evolution.

For all the DWDs formed at all metallicities in our \texttt{default} model, we simulate the complete temporal evolution of $f_\mathrm{e}$, $\dot{f_\mathrm{e}}$, and $\Mchirp$, starting from their initial orbital properties predicted by \cosmic\ at the instant of DWD formation. We inject these numerical expressions in Eq.~\ref{eq:realistic} and compute the total fluxes received in each frequency bin using the standard SFRD from \citet{madau14}.

\begin{figure}
    \centering
    \includegraphics{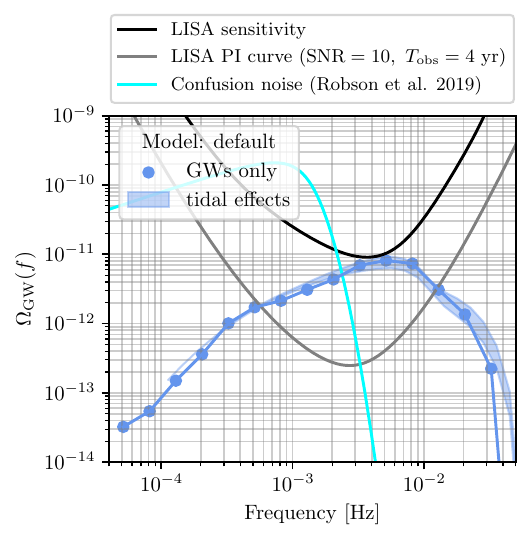}
    \caption{\AGWB\ when considering a more realistic treatment of DWD evolution, including episodes of mass transfer and tidal effects. The scatter points indicate the standard case with only GWs, with the \texttt{default} \cosmic\ model and SFRD from \citet{madau14}. The shaded area covers predictions from different assumptions about tidal effects.
    Additional curves represent the \textit{LISA} sensitivity, the PI curve, and the confusion noise from Galactic DWDs.
    }
\label{fig:tides}
\end{figure}

We show in Fig.~\ref{fig:tides} the \AGWB\ obtained in the standard case assuming evolution through GW emission only (blue points), as well as the uncertainty in the energy density spectrum associated with uncertainties in the intensity of tidal effects (blue shaded area). We observe that the inclusion of DWD interactions modifies the SGWB energy density spectrum by a factor $\sim 0.8-1.5$ for supra-mHz frequencies. This is because these interactions occur when the two WDs are close enough to each other. \citet{toubiana24} find that the typical GW frequency at which the out-spiralling systems accumulate is around $1.3\ \mathrm{mHz}$. This roughly corresponds to the value of the observed GW frequency at which interactions start playing a noticeable role in our predictions of the SGWB.
For the highest frequencies ($f_\mathrm{r}\geq2\times10^{-2}\ \mathrm{Hz}$), the inclusion of tidal effects in the evolution of DWDs could increase their contribution to the SGWB by a factor up to $\sim3$.

The lower estimate of the predicted SGWB corresponds to the model `low spins, weak tides' of \citet{toubiana24}. Under these assumptions, some systems do not survive the first intense episode of mass transfer \citep[see e.g.][]{pakmor22}, and the amplitude of the SGWB can then be slightly reduced compared to the simplified case of DWD evolution through GW emission alone.
On the other hand, in the model `low spins, strong tides' of \citet{toubiana24} (upper edge of the blue shaded area) the tidal effects typically help to stabilise this first episode of mass transfer, and more systems survive. Under these assumptions, the amplitude of the SGWB is then pushed to slightly higher values in this frequency range due to the fact that some DWDs can contribute several times to the same frequency bin over the course of their evolution.

The inclusion of realistic interactions in the evolution of DWDs produces a noticeable impact on our predictions of the SGWB from these systems, but its effect appears to be less important than the model of cosmic star formation history or the treatment of the CE phase during the evolution of massive stellar binaries (see Fig.~\ref{fig:omega_SFRs_cosmicmodels}). For this reason, and in order to reduce the computational cost of our simulations, we have chosen not to include these realistic interactions in all the combinations of the different models used in this study. We believe, however, that it is important to bear in mind the contribution of these uncertainties for future analyses of the SGWB produced by extragalactic DWDs.

\begin{figure}  
    \centering
    \includegraphics{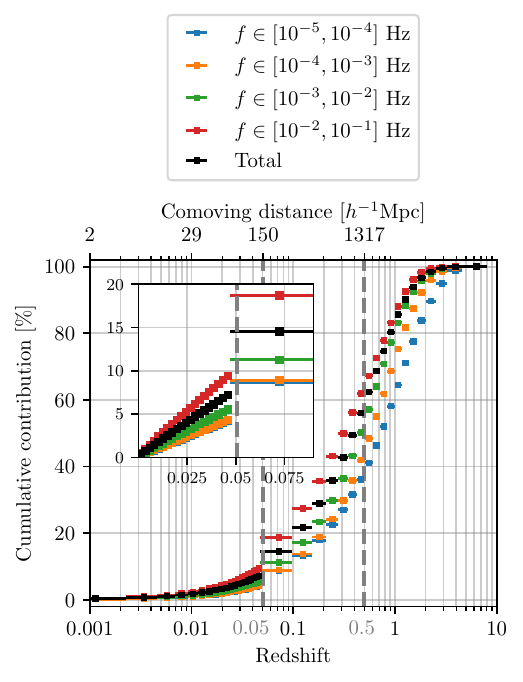}
    \caption{Cumulative contribution of extragalactic DWDs to the SGWB as a function of redshift, using our \texttt{default} population synthesis model and the SFRD from \citet{madau14}. The black dots represent the integrated contribution over the complete frequency range, and the coloured ones correspond to different specific frequency bands. The redshift marking the limit of the homogeneous isotropic Universe $z=0.05$ is indicated as the gray vertical dashed line, as is $z=0.5$.
    }
    \label{fig:anisotropy_contribution}
\end{figure}

\subsection{On the presence of anisotropies in the \AGWB}
\label{subsc:Anisotropies}

A practical exploration of anisotropies in the \AGWB\ could theoretically be carried out by sampling sources across a realistic spatial distribution of galaxies inside a cosmological volume. However, this approach would require the production of large catalogs of sources and to considerably modify the code developed for this study. We propose here a complementary analysis that uses the procedure presented in this study to place upper limits on the relative importance of anisotropies in the \AGWB.

In particular, we make the argument that if all the DWDs within a certain redshift range $[0,z_\mathrm{homogeneous}]$ contribute to $X\%$ of the total SGWB, and that the Universe is indeed homogeneous and isotropic beyond $z_\mathrm{homogeneous}$, then at most $X\%$ of this GW signal can carry some signature of an anisotropic origin.
We therefore consider the cumulative contribution of increasingly distant redshift shells to this SGWB as a tracer of the maximum impact that anisotropies in the distribution of galaxies could have on the GW signal. 

Following \cite{marinoni12}, we use a fiducial value of comoving distance $150\ h^{-1}\ \mathrm{Mpc}$ as the typical scale above which the Universe can be considered homogeneous and isotropic, which corresponds to a redshift $z_\mathrm{homogeneous}\simeq0.05$  \citep{planck20}.
We then split our initial first redshift bin in 20 new intervals between $0$ and $z_\mathrm{homogeneous}$ and complete our discretization with the other redshift bins previously used. We show in Fig.~\ref{fig:anisotropy_contribution} the cumulative contribution to the \AGWB\ as a function of redshift over the complete frequency spectrum (black curve), and for different specific frequency bands (coloured curves).
The very low contribution of the $z\leq0.05$ Universe in most frequency bins suggests that the hypothesis of homogeneous and isotropic Universe applied to derive Eq.~\ref{eq:integral_Om_AGWB} remains valid, at least to a few percent.

We find that the presence of anisotropies in the \AGWB\ would represent a greater fraction at the highest frequencies, with around $10\%$ of the signal in the range $[10^{-2},10^{-1}]\ \mathrm{Hz}$ originating from sources at $z\leq0.05$, against less than $5\%$ in the range $[10^{-5},10^{-4}]\ \mathrm{Hz}$. This fraction increases to $\sim60\%$ for sources in the range $[10^{-2},10^{-1}]\ \mathrm{Hz}$ at $z\leq0.5$, and $\sim40\%$ in the range $[10^{-5},10^{-4}]\ \mathrm{Hz}$.
As the distant systems emitting GWs at high frequencies have their signals redshifted to lower frequencies when observed on or near Earth, the range of high observed frequencies is naturally populated by close-by systems. This explains why the cumulative contribution is always relatively higher in the observed frequency range $[10^{-2},10^{-1}]\ \mathrm{Hz}$ compared to the other frequency ranges, although it is less important in absolute energy density. 

We remind the reader that the values indicated here represent  the relative contribution to the total signal in each frequency band. Indeed, as the amplitude of the \AGWB\ is predicted to be quite faint below $10^{-4}\ \mathrm{Hz}$ and to drop rapidly above $10^{-2}\ \mathrm{Hz}$, \textit{LISA} will most likely not be sensitive at all to the signal in these ranges of frequency (see Fig.~\ref{fig:omega_SFRs_cosmicmodels}). Looking specifically at the frequency range $[10^{-3},10^{-2}]\ \mathrm{Hz}$, which corresponds to the region where \textit{LISA} will be most sensitive and where the \AGWB\ is predicted to be most important, we find that only $\sim6\%$ of the total signal would originate from sources located at $z\leq0.05$. Even if all the sources inside that volume were located at one unique position, their combined GW signal would still be completely dominated by the more distant sources in the rest of the Universe. Therefore, the results presented in Fig.~\ref{fig:anisotropy_contribution} provide an upper limit on the maximal contribution of anisotropies to the total \AGWB, as a function of the limit of the homogeneous and isotropic Universe one considers.

A similar analysis carried on with the \texttt{alpha4} \cosmic\ model shows overall similar trends, although the contribution of the most distant sources appears to be slightly less important in the \texttt{alpha4} model. This is because the DWDs formed through more efficient CE evolution typically have larger orbital separations when they form, and therefore take longer to evolve to the \textit{LISA} frequency band, hence emitting at typically lower redshifts compared to the \texttt{default} model.

Considering additional sources of anisotropies in the GW signal from the modulated foreground, including contributions from sources in the Milky Way \citep{boileau21, bartolo22} and in satellite galaxies \citep{roebber20, rieck24}, the presence of anisotropies in the \AGWB\ becomes even more negligible. We conclude that the small amplitude of potential anisotropies in this SGWB will most likely not be observable by \textit{LISA}.

\section{Discussion}\label{sc:Discussion}

\begin{figure}
    \centering
    \includegraphics{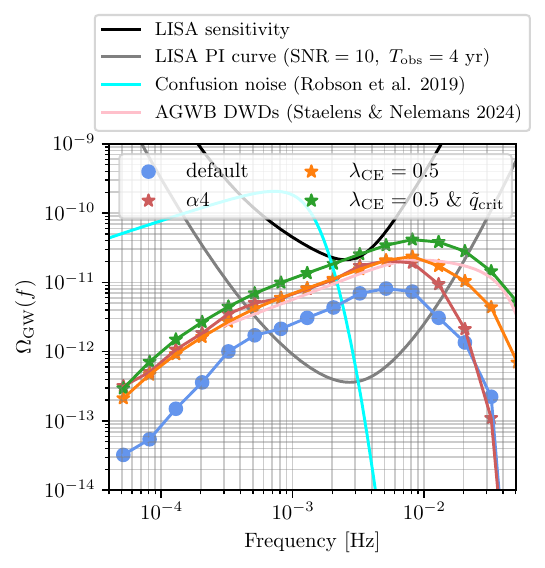}
    \caption{\AGWB\ with additional \cosmic\ models to mimic the different physical prescriptions applied in \texttt{SeBa}. The blue scatter points indicate the standard case, with our \texttt{default} \cosmic\ model and SFRD from \citet{madau14}. The red stars correspond to the \texttt{alpha4} model.
    The orange ones correspond to the SGWB obtained when fixing in \cosmic\ the CE efficiency parameter to $\alpha=4$ and the binding energy factor of a star to its envelope to a constant value $\lambda=0.5$. The green ones are obtained when using these same fixed values for $\alpha$ and $\lambda$, and modifying in \cosmic\ the criterion for mass transfer stability to resemble the one in \texttt{SeBa}.
    Additional curves represent the \textit{LISA} sensitivity, the PI curve, and the confusion noise from Galactic DWDs.
    The pink curve indicates the best-fit model of the \AGWB\ estimated in \citep{staelens24}.
    }
    \label{fig:Gijs}
\end{figure}

\subsection{Uncertainties in modeling the SGWB}
\label{subsc:SeBa}

In this work we have explored the impact of different uncertainties on the spectrum and amplitude of the \AGWB\ in the \textit{LISA} band. First of all, our results are globally consistent with other recent studies \citep{farmer03,staelens24,hofman24}. The spectrum from \citet{staelens24} lies in between two of our models, with different assumptions on the common envelope efficiency. 

One notable difference between our predictions and those of \citet{staelens24} and \citet{hofman24} lies in the high-frequency part of the energy density spectrum. Indeed, while they find models for backgrounds that extend up to $10^{-2}\ \mathrm{Hz}$ and start decreasing abruptly only for frequencies higher than $\sim3\times10^{-2}\ \mathrm{Hz}$, all our models are already falling-off at frequencies around $7\times10^{-3}\ \mathrm{Hz}$ (see Fig.~\ref{fig:omega_SFRs_cosmicmodels}). Looking in detail at the properties of the sources that contribute in this frequency range, we find that they are for the vast majority massive DWDs with tight orbits (with typical values of $\Mchirp\geq0.6\ \Msun$ and $a\leq3\ \Rsun$). The progenitors of these systems are stellar binaries formed with orbital separations already relatively low.

However, with our different \cosmic\ models, we barely manage to produce any of such massive and tight DWDs, even when evolving the exact same stellar progenitors as in \citet{staelens24}. Looking at the stellar evolution and various phases of binary interactions that are implemented within their code \texttt{SeBa} and within \cosmic, we were able to identify the main origin of the differences in the treatment of the CE evolution. Two factors play an important role here: first the criteria for the stability of mass transfer, which determines the stellar binaries that can evolve through stable mass transfer only until the formation of the two WDs, thus avoiding a potential merger during the CE phase; second the treatment of CE evolution itself.

In the version of the \texttt{SeBa} code used in \citet{staelens24} and \citet{hofman24}, they use a criterion based on the binary angular momentum to determine the stability of mass transfer, whereas in \cosmic\ we use a criterion based on the mass ratio. In practice, this makes it so that less systems experience CE phases in \texttt{SeBa} and more of these massive stellar binaries survive to eventually produce DWDs.
They also use a value of the binding energy factor of a star to its envelope to a constant value $\lambda=0.5$. This helps the survivability of the most massive systems. These same massive systems typically merge during the CE phase in \cosmic\, where we use a value of this binding energy factor that depends on stellar type \citep[following][]{claeys14}. 

To quantify the impact of these two different elements, we simulate two additional populations of DWDs with \cosmic, incorporating modifications to the \texttt{alpha4} model. In the first model, we set the binding energy factor to $\lambda=0.5$. In the second, we additionally modify the criterion for the stability of mass transfer inside \cosmic\ to mimic the one used in \texttt{SeBa}. The SGWB obtained for the standard \texttt{default} model, for the \texttt{alpha4} one, and for these two additional models are shown in Fig.~\ref{fig:Gijs}. 

We observe a clear shift of the high-frequency part of the SGWB with these different prescriptions of CE evolution. The second model (green scatter points), which incorporates both the modification for the onset of unstable mass transfer and for the treatment of CE itself, gives results that are in much better agreement with the predictions presented in \citet{staelens24} (pink curve) in the high-frequency regime.
In particular, the location of the typical frequency at which the signal starts to drop, referred to here as the knee, increases towards higher frequencies in these two additional models where the evolution of the CE is treated in a similar way to \texttt{SeBa}. 

The high frequency regime will also be the most affected by tidal torques and/or mass transfer between the DWDs. Following the \citet{toubiana24} model, we find up to a factor 2 variation in amplitude at the highest frequencies. While the high frequency regime appears to be the most interesting to study the effects of binary interaction, it is also the least likely to be detected, unless the signal amplitude is higher than predicted or the population of tight systems is more abundant than assumed. Given the high SNRs predicted, it may possible to detect signatures of tidal interactions or mass transfer effects in the observed signal. 

\subsection{Implications of the knee frequency}

Beyond the overall detectability of the \AGWB, the knee frequency provides particularly valuable astrophysical information. This knee frequency is primarily determined by the existence of very massive DWDs with extremely low orbital separations. 
In addition to the natural scaling of orbital frequency with the mass of the binary components, white dwarfs have the particularity of seeing their radius decrease as their mass increases \citep[see e.g.][]{kawaler97}. This implies that more massive DWDs can reach smaller orbital separations before one starts filling its Roche lobe, hence increasing again the maximum frequency at which they can emit GWs before their merger. To contribute to the high frequency part of the energy density power spectrum, these binaries must also exist at low redsfhits (otherwise their received frequency would be shifted to smaller values).
The location of the frequency knee in the \AGWB\ is therefore based on the possibility and efficiency of binary stellar evolution to produce two massive WDs very close to each other.

Models with more efficient envelope ejection (higher $\alpha$ or $\lambda$ parameters) tend to produce wider post-CE separations, but also allow some binaries to survive the CE evolution phase, whereas they would merge during this phase in the case of less efficient energy transfer to the envelope. This explains why the \AGWB\ predicted with our \texttt{alpha4} \cosmic\ model extends to slightly higher frequencies than our \texttt{default} model (see Fig.~\ref{fig:omega_SFRs_cosmicmodels} and Table~\ref{tab:appendix_parameters}).

This feature is explored in our comparison with the population synthesis code \texttt{SeBa} in the previous subsection (see e.g. the model $\lambda_\mathrm{CE}=0.5\ \&\ \tilde{q}_\mathrm{crit}$ and the results of \citet{staelens24} presented in Fig.~\ref{fig:Gijs}).

Therefore, the precise location of this spectral break encodes direct constraints on CE physics, which remains one of the largest uncertainties in binary stellar evolution.
Observing the knee frequency with LISA would thus offer an opportunity to constrain the physical prescriptions used for CE evolution and improve the predictions of population  synthesis models. Future analyses should therefore focus on precise modelling and measurement of the knee frequency, together with the overall amplitude, to enhance our understanding of binary evolution physics using the \AGWB\ observed by LISA.

\section{Conclusions}\label{sc:Conclusion}

In this study, we have presented new predictions for the SGWB produced by extragalactic DWDs, based on population synthesis models using COSMIC and three different SFRs. The amplitude and shape of the background strongly depend on key astrophysical assumptions, especially binary evolution physics and the SFR beyond redshift $z \sim 2$.

Our main findings can be summarised as follows:
\begin{itemize}
    \item The signal is detectable by \textit{LISA} across all scenarios explored, with SNRs ranging from \( \sim 100\) to \( \sim 400\) for conservative models ans exceeding \( \sim 1000\) for optimistic models (e.g., \texttt{Strolger} + \texttt{alpha4}). For 10-year integration, the SNR can reach up to \(\sim 2400\), indicating that the extragalatic DWD background will be robustly observed by LISA under all astrophysical assumption considered here.
    
    \item The dominant uncertainty on $\Omega_{\mathrm{GW}}$ arises from the cosmic SFR (factor $\sim 6$–7), with additional variation from binary physics (factor $\sim 2$–3).
    
    \item Tidal effects and post-formation mass transfer introduce a further modulation above 5~mHz, at the level of a factor $\sim 2$–3. These remain subdominant compared to population-wide assumptions.
    \item Our findings are in line with previous studies in terms of amplitude of the signal, and power law behavior at low frequency. We confirm the variations based on star formation model and binary evolution models.
    
    \item A spectral break occurs around 7~mHz in our models, in contrast to earlier predictions near 20~mHz. This shift reflects differences in the formation and survival of compact, massive systems following mass transfer.
    
    \item The signal is largely isotropic, with contributions from the nearby universe ($z < 0.05$) at the percent level, insufficient for detection of anisotropies.

\end{itemize}

Reducing model uncertainty requires improved treatment of binary evolution, particularly during the common-envelope phase. Multi-messenger comparisons, such as those discussed in \citet{vanzeist25}, offer promising avenues to calibrate population synthesis outputs using electromagnetic constraints. Further cross-validation between independent codes (e.g., BPASS~\citep{stanway23, byrne25}) will also help identify robust features in SGWB predictions.
Interpreting this detection will require careful modelling of foregrounds and explicit inclusion of the \AGWB\ in future LISA data analysis pipelines. A more detailed analysis of its detectability and implications will be the subject of future works.

These results also underline that no reliable estimate of the detectability of cosmological SGWBs can be made without first accounting for the extragalactic DWD background. Its amplitude and spectral shape make it a dominant component in the \textit{LISA} band. Accurate modelling of this signal is therefore essential to define the astrophysical foreground and to interpret any potential cosmological contribution. Ignoring it would lead to overly optimistic forecasts and biased conclusions \citep{boileau21, boileau22, boileau23}. The \AGWB\ represents a dominant signal of SGWB in the \textit{LISA} band and will severely limit the detectability of cosmological SGWBs. It must therefore be explicitly included in future simulations and in the data analysis pipelines developed within the \textit{LISA} Science Ground Segment framework.

\begin{acknowledgements}
We sincerely thank the group of Gijs Nelemans for their valuable assistance and the constructive discussions that helped us better understand the differences between our results. We would also like to thank Arianna Renzini and Riccardo Buscicchio for their priceless comments and advice. Their insights were instrumental in refining our analysis and interpretations. G.B. thanks the Centre national d'études spatiales (CNES) for support for this research.  This project has received financial support from the CNRS through the MITI interdisciplinary programs through its exploratory research program. A.L., T.B. and N.C. acknowledge support by the French Agence Nationale de la Recherche. 
T.B. and A.T. are supported by ERC Starting Grant No.~945155--GWmining, Cariplo Foundation Grant No.~2021-0555, MUR PRIN Grant No.~2022-Z9X4XS, MUR Grant ``Progetto Dipartimenti di Eccellenza 2023-2027'' (BiCoQ), and the ICSC National Research Centre funded by NextGenerationEU. A.T. is supported by MUR Young Researchers Grant No. SOE2024-0000125.
\end{acknowledgements}

\bibliographystyle{aa}
\bibliography{references}
 
\begin{appendix}
\section{Comparison with previous studies}\label{app:comparison}

Here we present numbers and plots allowing direct comparison with \citet{staelens24}, and provide additional figures detailing the contributions of different types of DWDs.

\begin{figure}
    \centering
    \includegraphics{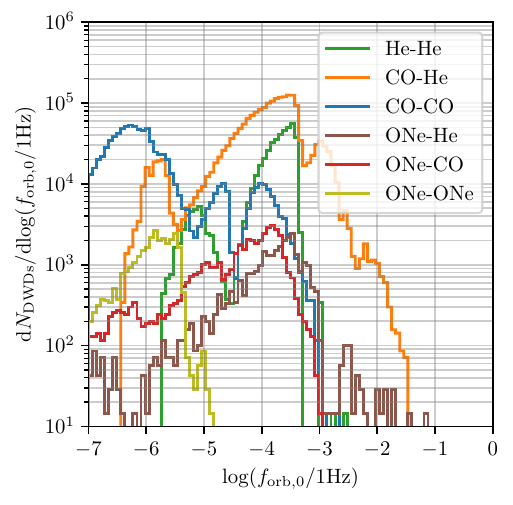}
    \caption{Distribution of orbital frequencies at the instant of formation $f_\mathrm{orb,0}$ of the DWDs produced with our \texttt{default} \cosmic\ model at Solar metallicity. This set of DWDs contains only those that have passed the low frequency cut-off and could potentially end up in the \textit{LISA} band. Each coloured curve represents a specific type of DWDs.
    }
\label{fig:Dist_Freq_TypeDWD}
\end{figure}

\begin{figure}
    \centering
    \includegraphics{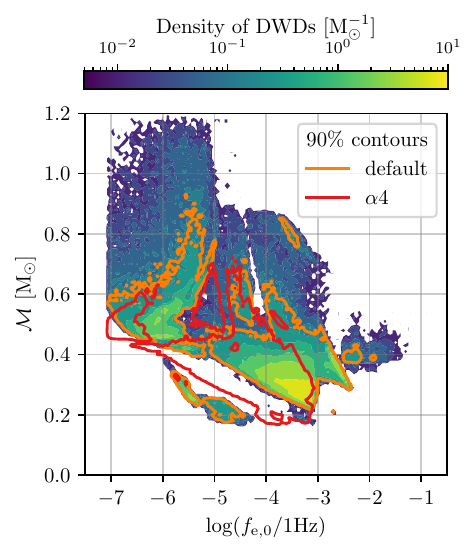}
    \caption{Density plot of the chirp mass $\Mchirp$ and GW frequency at the time of formation $f_\mathrm{e,0}$ of all the DWDs produced by our \texttt{default} \cosmic\ model at solar metallicity (colormap). Solid lines show the contours containing 90\% of the DWDs in the parameter space ($f_\mathrm{e,0}$, $\Mchirp$) for this \texttt{default} model (orange) as well as for the \texttt{alpha4} model (red).}
\label{fig:DensityMap_DWD_Fe_Mc}
\end{figure}

\subsection{Physical properties of DWDs at formation and formation efficiency}
\label{app:DWDs}

Fig.~\ref{fig:Dist_Freq_TypeDWD} shows the distribution of orbital frequencies at the instant of formation for the DWDs obtained with our \texttt{default} \cosmic\ model.
CO-CO pairs form from originally wide stellar binaries that do not undergo envelope stripping before He begins to burn in the core, and generally result in wide DWD systems with therefore fairly low orbital frequencies. 
He-He pairs, on the other hand, come from two lower mass stars in a tight binary where the two stars have had their envelopes removed by binary interactions at some point in their evolution. This results in DWD systems with typically higher initial orbital frequencies. 
CO-He systems are formed through a combination of these two phenomena.
The systems containing at least one ONe WD originate from more massive stars and are naturally less common. As only the widest of these massive binaries can evolve without undergoing stellar mergers, ONe-ONe pairs are typically formed with large orbital separations, meaning low initial orbital frequencies, and most of these systems emit so few GWs that they will never actually enter the \textit{LISA} band.

\begin{figure}
    \centering
    \includegraphics{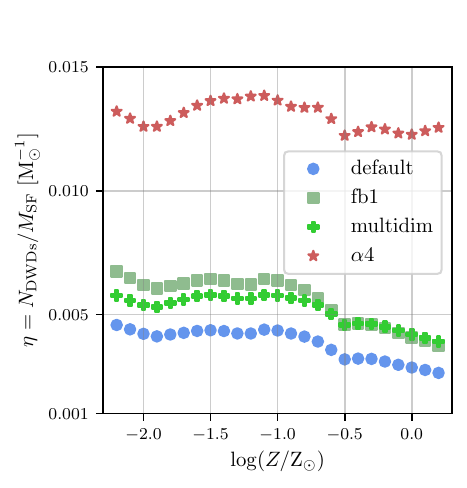}
    \caption{Formation efficiency of DWDs as a function of metallicity for the four \cosmic\ models considered in this study. At each metallicity, $\eta$ is computed as the number of DWD sytems formed above the low frequency cut-off divided by the total mass of the stellar popultion simulated in \cosmic.}
\label{fig:etaDWD}
\end{figure}

Fig.~\ref{fig:DensityMap_DWD_Fe_Mc} presents the distribution of all DWD systems at formation in the parameter space ($f_\mathrm{e,0}$,$\Mchirp$), where $f_\mathrm{e,0}$ is the emission frequency of GWs at the instant of DWD formation and $\Mchirp$ is the chirp mass. This figure can be directly compared with Figure 1 in \citet{staelens24}.  As the first three models \texttt{default}, \texttt{fb1}, and \texttt{multidim}, use the exact same physics to describe binary evolution but differ only in the properties of their initial stellar populations, they predict DWD systems with overall similar masses and orbital properties. On the other hand, the \cosmic\ model \texttt{$\alpha$4} with a higher CE efficiency parameter results in a population of DWDs with typically smaller GW frequencies at formation.The peak of the distribution lies around $f_\mathrm{e,0}\lesssim10^{-3}\ \mathrm{Hz}$ and $\Mchirp\sim0.3\ \Msun$, which correspond to the typical masses and frequencies at which CO-He and He-He pairs are formed (see Fig.~\ref{fig:Dist_Freq_TypeDWD}).

In order to consistently compare DWD formation across our different \cosmic\ models, we consider the DWD formation efficiency defined as
\begin{equation}
    \eta(Z)=\frac{N_\mathrm{DWDs}(Z)}{M_\mathrm{SF}(Z)} ,
\end{equation}
where $N_\mathrm{DWDs}(Z)$ is the number of DWDs formed from the evolution of the stellar population initialised with metallicity $Z$, and $M_\mathrm{SF}(Z)$ is the total initial mass of the stellar population, including both single and binary stars.
We show in Fig.~\ref{fig:etaDWD} the evolution of this DWD formation efficiency $\eta$ as a function of metallicity for our four \cosmic\ models. As WDs stem from only moderately massive stars, where mass-loss through stellar winds is relatively less important, the impact of metallicity on the formation of DWDs is expected to be only a minor effect. We observe that for all models, there is a factor of at most $\sim1.5$ between $\eta(0.01 \Zsun)$ and $\eta(\Zsun)$, which confirms the moderate influence of metallicity on DWD formation.
The \texttt{default} model is consistently the least efficient in forming DWDs, whereas the models \texttt{$f_\mathrm{b}=1$} and \texttt{multidim} agree for the most part across all metallicities. The model \texttt{$\alpha$4} predicts a DWD formation efficiency roughly 3 times that of the other three models, and shows no clear sign of decreasing with metallicity.

\begin{figure*}
    \centering
    \includegraphics{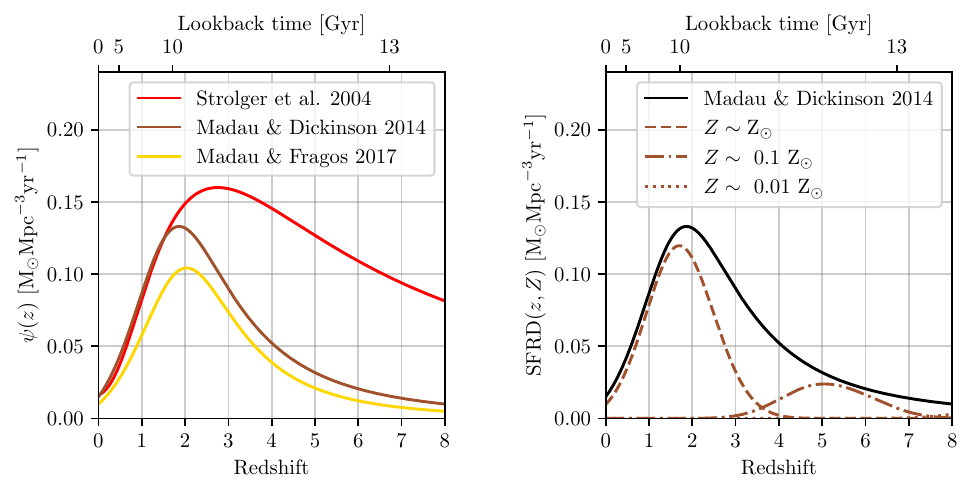}
    \caption{\textbf{Left:} SFRD in the Universe as a function of redshift. Different colours indicate different estimates based on different assumptions on the initial mass function of stellar populations or on extinction correction \citep{strolger04,madau14,madau17}.
    \textbf{Right:} Metallicity-specific SFRD as a function of redshift for different values of metallicity, computed as the combination of the SFRD from \citet{madau14} with the metallicity distribution of \citet{neijssel19}. The dashed, dot-dashed, and dotted lines show the metallicity-specific SFRD around metallicity $\Zsun$, $0.1\times\Zsun$, and $0.01\times\Zsun$ respectively.
    }
\label{fig:SFRs}
\end{figure*}

\subsection{Metallicity-dependent star formation history}
\label{app:SFR}

\begin{table}
\caption{Best-fit parameters of the three different SFRD models considered in this study.}
\label{tab:sfr_models}
\centering
\begin{tabular}{lcccc}
\hline\hline
\textbf{Model \(\psi(z)\)} & \textbf{\(a\)} & \textbf{\(b\)} & \textbf{\(c\)} & \textbf{\(d\)} \\
\hline
Madau \& Dickinson (2014) & 0.015 & 2.7 & 2.9 & 5.6 \\
Madau \& Fragos (2017)    & 0.010 & 2.6 & 3.2 & 6.2 \\
\hline
\noalign{\smallskip}
\textbf{Model \(\widetilde{\psi}(t)\)} & \textbf{\(\tilde{a}\)} & \textbf{\(\tilde{b}\)} & \textbf{\(\tilde{c}\)} & \textbf{\(\tilde{d}\)} \\
\hline
Strolger et al. (2004) & 0.182 & 1.26 & 1.865 & 0.071 \\
\hline
\end{tabular}
\tablefoot{Parameters correspond to the analytical forms defined in~\cite{strolger04} \cite{madau14}) and~\cite{madau17}, and show in \textbf{left} Fig~\ref{fig:SFRs}.)}
\end{table}

The left-hand panel of Fig.~\ref{fig:SFRs} shows the SFRDs according to these three different models in this paper. They all agree reasonably well at low redshifts ($z\lesssim2$) but the model from \citet{strolger04} finds substantially more star formation at higher redshifts.  The right-hand panel of Fig.~\ref{fig:SFRs} shows  the results of the combination of the SFRD from \citet{madau14} with the metallicity distribution function of \citet{neijssel19}. Under these assumptions, most of star formation takes place at $z\leq2$ and involves already metal-enriched stars (at $Z\sim\Zsun$), while the formation of stars at $Z\sim0.1\ \Zsun$ principally takes place at around $z\sim5$. The most metal-poor stars at $Z\sim0.01\ \Zsun$ are rare and form almost exclusively at the highest redshifts ($z\lesssim8$).

\subsection{Progenitor metallicity of the DWDs contributing to the SGWB}
\label{app:metallicity}

\begin{figure*}
    \centering
    \includegraphics{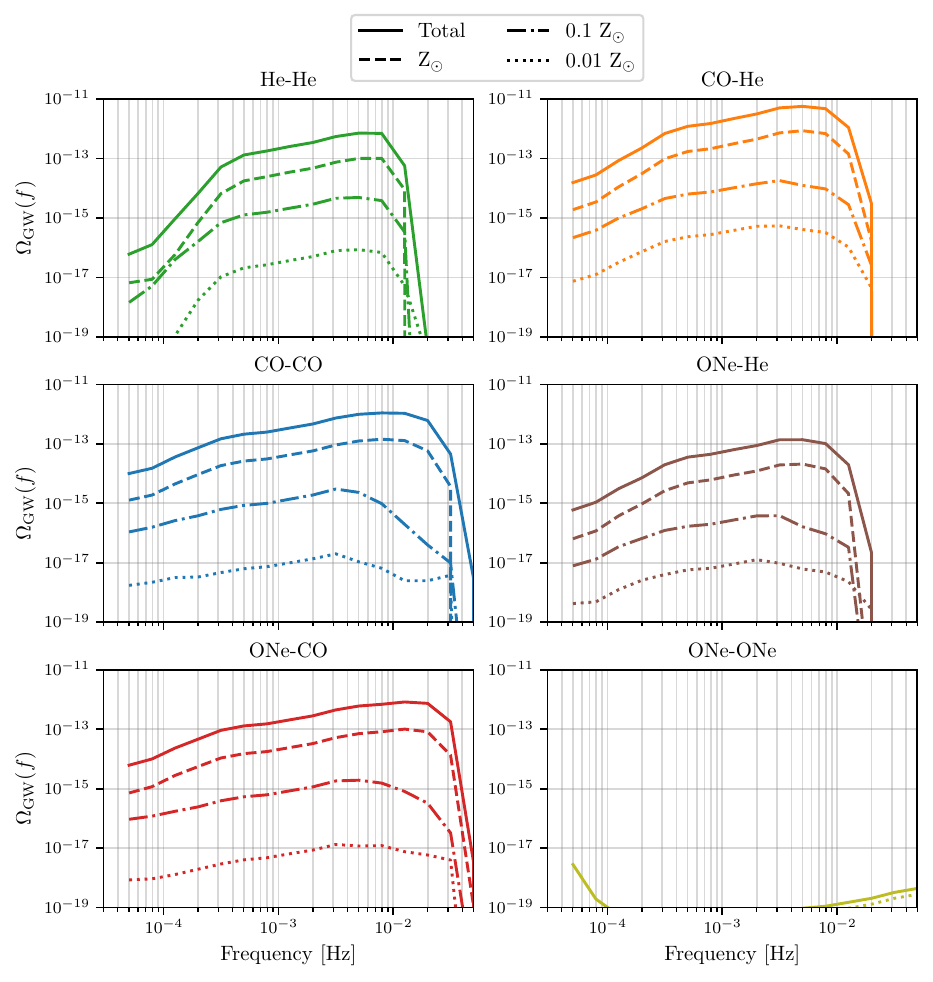}
    \caption{\AGWB\ for the various types of DWDs and for different metallicities of stellar progenitors. In each sub-figure, the solid coloured line shows the energy density spectrum of the total SGWB from this type of DWDs while the dashed, dot-dashed, and dotted lines show the contribution of DWDs formed by the evolution of stellar progenitors around metallicity $\Zsun$, $0.1\times\Zsun$, and $0.01\times\Zsun$ respectively. Coloured lines indicate the contributions from different DWD types. The plot highlights the overall SGWB with different line styles for each type of DWD binary: CO-CO binaries are shown in blue, CO-He binaries in orange, He-He binaries in green, ONe-CO binaries in red, ONe-He binaries in brown, and ONe-ONe binaries in olive.}
\label{fig:freq_vs_omega_gw}
\end{figure*}

Fig.~\ref{fig:freq_vs_omega_gw} presents the energy density spectrum for the different types of DWDs as a function of progenitor metallicity, using our \texttt{default} \cosmic\ model and the SFRD from \citet{madau14}. We show here only the contributions to the SGWB for stellar progenitors formed at metallicities at $\Zsun$ (dashed lines), $0.1\times\Zsun$ (dot-dashed lines), and $0.01\times\Zsun$ (dotted lines), but the total energy density spectrum, plotted as solid lines, are still estimated for all the systems formed in our \cosmic\ model from the 25 different metallicities simulated. For all types of DWDs, with the exception of ONe-ONe binaries whose total contribution is negligible at all frequencies, the SGWB is dominated by systems formed through the evolution of  binaries around $\Zsun$. This is directly inherited from our model of the metallicity-specific SFRD (see Eq.~\ref{eq:ZSSFR}), which predicts that most of star formation takes place at redshift $z\sim2$ in environments where the metallicity is already close to the Solar value (see Fig.~\ref{fig:SFRs}). Although systems formed at higher redshifts from metal-poor stellar progenitors may have more time to evolve and eventually enter the \textit{LISA} frequency band, there are just not enough of them due to the limited star formation at these redshifts.

\section{Estimating the SGWB Using a Smoothly Broken Power Law fit }\label{app:parameters}

This appendix gathers the results of the SGWB energy spectral density summarize on Fig.~\ref{fig:omega_SFRs_cosmicmodels}. We investigate whether the \AGWB\ can be effectively modeled by a simple analytical expression. For that, we adopt a smoothly broken power-law function and assess how well it fits our simulated SGWB across astrophysical synthesis scenarios or hypotheses. This model is motivated by the physical properties of the DWD population contributing to the \AGWB. At low frequencies, the background follows a power-law behavior characteristic of a superposition of binary inspirals. However, the finite size and physical constraints of white dwarfs prevent them from emitting efficiently at high frequencies, leading to a depletion of sources and thus a spectral turnover. This transition defines a break in the spectrum. The break is smooth rather than abrupt, as the underlying binary population exhibits a distribution of physical parameters such as masses and orbital separations.

We fit a smoothly broken power-law model to estimate the \AGWB\ across the different scenarios of SFRD and initial population parameters introduced in this paper. The model is defined as follows:

\begin{equation}
    \Omega_{\text{Model}}(f) = A \left( \frac{f}{f_b} \right) ^ {-\alpha_1}
       \left\{\frac{1}{2}  \left[1 + \left( \frac{f}{f_b}\right)^{1 / \Delta}\right]\right\}^{(\alpha_1 - \alpha_2) \Delta} ,
\end{equation}
where \(A\) is the amplitude, \(f_\mathrm{b}\) is the frequency location of the knee, \(\alpha_1\) and \(\alpha_2\) are the spectral indices before and after this frequency, and \(\delta\) controls the smoothness of the transition. The implementation of this model was facilitated using the smoothly broken power law from the \texttt{astropy} library \citep{astropy:2018, astropy:2022}. 

In our analysis, the spectral fitting of the \AGWB\ is carried out using the \texttt{dynesty} library \citep{dynesty:2020}, which implements a nested-sampling Markov chain Monte Carlo (MCMC) algorithm. This Bayesian approach yields posterior distributions for the model parameters, their uncertainties, and the model evidence for possible comparison. The likelihood is defined as the sum of squared differences between the model and the simulated \AGWB\ spectrum, weighted by the standard deviation of data accross the fitted region. 

This methodology follows closely the approach developed in previous studies of the SGWB for \textit{LISA},  which employed similar MCMC-based strategies for spectral parameter estimation~\citep{boileau21,boileau22}. In particular, \citet{boileau21} introduced an adaptive MCMC algorithm to analyze simulated \textit{LISA} data, demonstrating the effectiveness of Bayesian inference in separating astrophysical and cosmological background components. Building on this, \citet{boileau22} extended the Bayesian MCMC framework to assess the detectability of a cosmologically-generated background in the presence of a foreground, highlighting the relevance of such techniques for SGWB analyses. By adopting a nested-sampling MCMC approach in this work, we ensure methodological continuity with these studies, which strengthens the reliability of our \AGWB\ spectral estimation and enables direct comparison with previous results in the \textit{LISA} context.

\begin{figure}[htbp]
    \centering
    \includegraphics[width=0.5\textwidth]{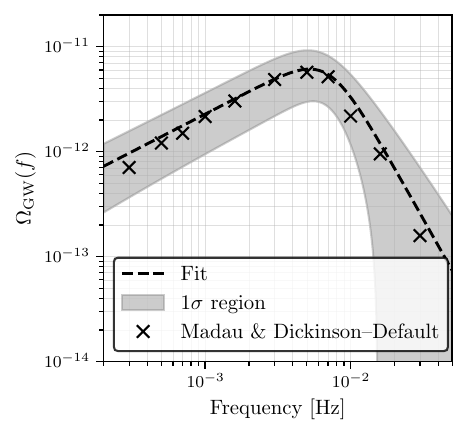}
    \caption{Total gravitational-wave background spectrum computed using the \texttt{default} model for the DWD population and the Madau \& Dickinson star formation history. The solid line represents the fit, and the shaded area indicates the $1\sigma$ uncertainty region.}
    \label{fig:vanhaaften_madau_fit}
\end{figure}

The parameters of the fit for each SFRD and initial population case are summarised in Table~\ref{tab:appendix_parameters}, and we show a comparison between the fit and the true signal for the \texttt{default} model in Fig.~\ref{fig:vanhaaften_madau_fit}. For each fit, we initialise a range of plausible parameters informed by the physical constraints of the \AGWB. The nested sampling approach ensured accurate estimation of posterior distributions. 

\begin{table*}[ht]

\centering
\caption{Fitted parameters for different initial populations. The uncertainties correspond to 1$\sigma$ errors.}
\begin{tabular}{llllll}
\hline

\multicolumn{2}{c}{\textbf{Parameters}} & \multicolumn{1}{c}{\( A \)} & \multicolumn{1}{c}{\( f_b \, [\mathrm{Hz}] \)} & \multicolumn{1}{c}{\textbf{\(\alpha_1, \alpha_2\)}} & \multicolumn{1}{c}{\textbf{\(\Delta\)}} \\

\hline
\hline

\multicolumn{6}{c}{\textbf{Madau Dickinson}} \\ \hline
\multirow{4}{*}{
   \rotatebox{90}{\shortstack{\textbf{Stellar} \\ \textbf{Model}}}}

& \texttt{default} & \(5 \pm 3 \times 10^{-12}\) & \(0.007 \pm 0.001\) & \(-0.72 \pm 0.09, 2.43 \pm 1.22\) & \(0.24 \pm 0.07\) \\ \cline{2-6}
& \texttt{$\alpha$4} & \(10 \pm 6 \times 10^{-12}\) & \(0.007 \pm 0.001\) & \(-0.69 \pm 0.09, 2.74 \pm 1.43\) & \(0.26 \pm 0.08\) \\ \cline{2-6}
& \texttt{fb1} & \(7 \pm 1 \times 10^{-12}\) & \(0.007 \pm 0.001\) & \(-0.74 \pm 0.08, 2.31 \pm 1.05\) & \(0.23 \pm 0.06\) \\ \cline{2-6}
& \texttt{multidim} & \(6 \pm 2 \times 10^{-12}\) & \(0.007 \pm 0.001\) & \(-0.73 \pm 0.08, 2.41 \pm 1.20\) & \(0.24 \pm 0.07\) \\ \hline

 \hline
\multicolumn{6}{c}{\textbf{Madau Fragos}} \\ \hline

\multirow{4}{*}{
   \rotatebox{90}{\shortstack{\textbf{Stellar} \\ \textbf{Model}}}}
   
& \texttt{default} & \(4 \pm 2 \times 10^{-12}\) & \(0.007 \pm 0.001\) & \(-0.72 \pm 0.08, 2.41 \pm 1.18\) & \(0.24 \pm 0.07\) \\ \cline{2-6}
& \texttt{$\alpha$4} & \(6 \pm 2 \times 10^{-12}\) & \(0.008 \pm 0.001\) & \(-0.76 \pm 0.07, 4.20 \pm 0.73\) & \(0.22 \pm 0.05\) \\ \cline{2-6}
& \texttt{fb1} & \(4 \pm 2 \times 10^{-12}\) & \(0.007 \pm 0.001\) & \(-0.72 \pm 0.09, 2.45 \pm 1.23\) & \(0.24 \pm 0.07\) \\ \cline{2-6}
& \texttt{multidim} & \(6 \pm 1 \times 10^{-12}\) & \(0.007 \pm 0.001\) & \(-0.75 \pm 0.08, 2.77 \pm 0.95\) & \(0.22 \pm 0.06\) \\ \hline

\hline
\multicolumn{6}{c}{\textbf{Strolger}} \\ \hline

\multirow{4}{*}{
   \rotatebox{90}{\shortstack{\textbf{Stellar} \\ \textbf{Model}}}}
   
& \texttt{default} & \(20 \pm 6 \times 10^{-12}\) & \(0.007 \pm 0.001\) & \(-0.75 \pm 0.08, 3.71 \pm 0.76\) & \(0.22 \pm 0.05\) \\ \cline{2-6}
& \texttt{$\alpha$4} & \(30 \pm 9 \times 10^{-12}\) & \(0.010 \pm 0.001\) & \(-0.69 \pm 0.03, 5.11 \pm 0.51\) & \(0.31 \pm 0.03\) \\ \cline{2-6}
& \texttt{fb1} & \(30 \pm 10 \times 10^{-12}\) & \(0.007 \pm 0.001\) & \(-0.75 \pm 0.08, 4.23 \pm 0.77\) & \(0.22 \pm 0.05\) \\ \cline{2-6}
& \texttt{multidim} & \(20 \pm 9 \times 10^{-12}\) & \(0.007 \pm 0.001\) & \(-0.75 \pm 0.08, 3.27 \pm 0.92\) & \(0.22 \pm 0.06\) \\ \hline

\end{tabular}
\label{tab:appendix_parameters}
\end{table*}

\end{appendix}

\end{document}